\documentclass{WileyMSP-template}
\usepackage[T1]{fontenc} 
\usepackage[dvipsnames]{xcolor} 
\colorlet{cite}{LimeGreen!50!Green}
\usepackage[utf8]{inputenc}
\usepackage{soul}
\usepackage{mathrsfs} 
\usepackage{tikz}  
 \usepackage{hyperref}
\usetikzlibrary{arrows,positioning}
\tikzset{ 
  baseline=-2.3pt,
  text height=1.5ex, text depth=0.25ex,
  >=stealth,
  node distance=2cm,
  mid/.style={fill=white,inner sep=2.5pt},
}
\usepackage{lmodern}
\usepackage{tikz-cd}
\usepackage{amsthm, amssymb, amsfonts,amsmath}
\usepackage{textcomp} 
\usepackage{multirow}
\usepackage{cancel}
\usepackage{diagbox}
\usepackage{makecell}
\theoremstyle{plain}	

\newtheorem*{theorem*}{Theorem}
\theoremstyle{definition}

\newtheorem*{conjecture*}{Conjecture}
\theoremstyle{remark}

\newtheorem*{lemma*}{Lemma}

\begin{document}
\emergencystretch 3em
\makeatletter
\let\old@fpheader\@fpheader
\pagestyle{fancy}
\rhead{\includegraphics[width=2.5cm]{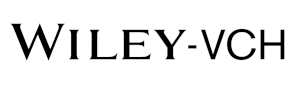}}
\title{Non-Invertible Symmetries in Compactified Supergravities}

\maketitle


\author{Fabián Caro-Pérez}
\author{María Pilar García del Moral}
\author{Alvaro Restuccia}



\begin{affiliations}
F. Caro-Pérez\\
Departamento de Física, Universidad de Antofagasta,\\ Universidad de Antofagasta, Campus Coloso, Aptdo 02800, Antofagasta, Chile\\
fabian.caro.perez@ua.cl\\
 Prof. M. P. García del Moral\\
Área de Física, Departamento de Química e Instituto de Computaci\'on  Cient\'ifica (SCRIUR),Universidad de la Rioja,
 C/ Madre de Dios 53, Logroño 26006, La Rioja, España.\\
m-pilar.garciam@unirioja.es

 Prof. A. Restuccia\\
 Departamento de Física, Universidad de Antofagasta,\\ Universidad de Antofagasta, Campus Coloso, Aptdo 02800, Antofagasta, Chile\\
 alvaro.restuccia@uantof.cl
\end{affiliations}


\begin{abstract}
We study the \textit{Kaluza--Klein} descent of non-invertible higher-form symmetry
defects from eleven-dimensional Supergravity to Type IIA Supergravity. Starting from
the eleven-dimensional construction of non-invertible Supergravity defects,
we show that the full defect, including its auxiliary topological sector, admits a pushforward along the M-theory circle in the zero-mode Supergravity regime. The seven-dimensional \textit{Chern--Simons}-like auxiliary theory
descends to a six-dimensional \(BF\)-type sector. We also show that the compactification
of the eleven-dimensional \textit{Bianchi} sector splits into an invertible
\(H_{[3]}\)-sector and a twisted non-invertible \(\widetilde F_{[4]}\)-sector, controlled by
\(d\widetilde F_{[4]}+H_{[3]}\wedge F_{[2]}=0\). The resulting Type IIA defect algebra
contains both an invertible Picard subgroup and non-invertible \(BF\)-dressed defects,
with charged probes identified through the standard M-theory/Type IIA brane
dictionary.
\end{abstract}

\section{Introduction}
\noindent
Symmetries have constituted during a long time a foundational principle in quantum field theory. In the standard formulation, a continuous global symmetry is defined by a group acting on local operators and, in the absence of anomalies, by a conserved Noether current. The corresponding conserved charge acts on the Hilbert space as an invertible operator. In this framework, the composition of two symmetry transformations produces another symmetry transformation, each element has an inverse, and the resulting algebraic structure is a group.

\noindent
Over the past decade, the concept has been significantly expanded. The primary progress is the introduction of generalized global symmetries \cite{Gaiotto:2014kfa}, \cite{Bhardwaj:2023kri}, \cite{Cordova:2022ruw}. A \(q\)-form global symmetry acts not on local operators, but rather on extended operators of dimension \(q\). The associated charged objects include lines, surfaces, or higher-dimensional defects, while the symmetry generators are topological operators supported on codimension-\((q+1)\) manifolds. In a \(d\)-dimensional quantum field theory, a \(q\)-form symmetry\footnote{Here, a $q$-symmetry, for simplicity, will be denoted as $\mathrm{HFS}(q)$.} is generated by their topological defects.
\begin{align}
    U_g(\Sigma_{d-q-1}),
\end{align}
where \(\Sigma_{d-q-1}\) is a closed codimension-\((q+1)\) cycle and \(g\) is an element of the symmetry group. The topological nature of \(U_g\) means that correlation functions are invariant under smooth deformations of \(\Sigma_{d-q-1}\), provided that the defect does not cross charged operators. Charged operators are \(q\)-dimensional operators \(W_q(\mathcal{C}_q)\) supported
on \(q\)-cycles \(\mathcal{C}_q\), and their charge is detected by linking with the topological symmetry operator:
\begin{align}
    U_g(\Sigma_{d-q-1})\, W_q(\mathcal{C}_q)={}&\mathcal{X}_g(W_q)^{\operatorname{Link}(\Sigma_{d-q-1},\mathcal{C}_q)}W_q(\mathcal{C}_q).
\end{align}
Where $\mathcal{X}_g(W_q)^{\operatorname{Link}(\Sigma_{d-q-1},\mathcal{C}_q)}$ is a phase\footnote{In general $\mathcal{X}_g(W_q)^{\operatorname{Link}(\Sigma_{d-q-1}, \mathcal{C}_q)}\sim\exp (2\pi i Q~{\operatorname{Link}(\Sigma_{d-q-1},\mathcal{C}_q)})$, with $Q$ charge of $W_q$ and $\operatorname{Link}(\Sigma_{d-q-1}, \mathcal{C}_q)$ is the linking number \cite{Gaiotto:2014kfa}.}. This formulation unifies many familiar structures in gauge theories and string theory.
Wilson lines are charged under electric one-form symmetries, while \textit{'t~Hooft}
lines are charged under magnetic one-form symmetries. More generally, stringlike
operators are charged under two-form symmetries, and higher-dimensional brane
operators furnish charged objects for higher-form symmetries. In particular, in
M-theory the M2- and M5-brane operators naturally appear as charged objects for
\(\mathrm{HFS}(3)\) and \(\mathrm{HFS}(6)\), respectively
\cite{Gaiotto:2014kfa,Fernandez-Melgarejo:2024ffg,Garcia-Valdecasas:2023mis}.

\noindent
The second, more recent, extension is the notion of non-invertible symmetry \cite{Shao:2023gho}, \cite{Schafer-Nameki:2023jdn}, \cite{Heckman:2024obe}, \cite{Heidenreich:2020pkc}. The key step is to identify symmetries not necessarily with groups, but with topological defects. Once this point of view is adopted, there is no fundamental reason why every topological defect implementing a selection rule or Ward identity should admit an inverse under fusion. In a general quantum field theory, topological defects may obey fusion rules of the form
\begin{align}
    \mathcal{D}_a \otimes \mathcal{D}_b=\bigoplus_c N_{ab}^{\;\;c}\,\mathcal{D}_c,
\end{align}
with non-negative integer fusion coefficients \(N_{ab}^{\;\;c}\in \mathbb{N}\). More generally, especially in higher spacetime dimensions, the outcome of fusing two defects do not need to be just a direct sum of simple defects. It may also contain a residual lower-dimensional topological field theory supported on the fusion locus. Thus, the fusion product of defects defines a categorical structure which extends the ordinary notion of group multiplication. A defect is called non-invertible when it does not admit an inverse under fusion. Namely, if for some defect \(\mathcal{D}_a\) there is no defect \(\mathcal{D}_a^{-1}\) such that
\begin{align}
    \mathcal{D}_a \otimes \mathcal{D}_a^{-1}={}&\mathbf{1},
\end{align}
then \(\mathcal{D}_a\) generates a non-invertible symmetry
\(\widehat{\mathrm{HFS}}(q)\)\footnote{We will generally refer to this non-invertible
symmetry of degree \(q\) as \(\widehat{\mathrm{HFS}}(q)\).}. The algebraic structure is no
longer a group. It is instead described by a fusion category, a higher fusion category, or
more generally by a topological symmetry structure whose defects encode fusion rules,
anomalies and possible obstructions to gauging
\cite{Shao:2023gho,Thorngren:2019iar,Gagliano:2024off,GarciaEtxebarria:2022vzq}.

\noindent
The canonical low-dimensional example is the \textit{Kramers--Wannier} duality defect of
the two-dimensional Ising model \cite{Frohlich:2004ef}. This defect is topological at the
critical point, but it is not invertible. Its self-fusion produces a sum of invertible symmetry
lines rather than the identity alone:
\begin{align}
    \mathcal{N}\otimes \mathcal{N}
    =
    \mathbf{1}\oplus \eta,
\end{align}
where \(\eta\) is the ordinary \(\mathbb{Z}_2\) symmetry line, with $\eta^2=\mathbf{1}$. This elementary fusion rule captures the essential conceptual shift. A non-invertible symmetry may still impose exact selection rules and dynamical constraints, although it does not act as a group on the Hilbert space \(\mathscr{H}\).

\noindent
In higher dimensions, the \(\widehat{\mathrm{HFS}}(q)\) can be constructed by several related
mechanisms. One of the most important is half-gauging
\cite{Schafer-Nameki:2023jdn,Choi:2022jqy}. Suppose a theory is invariant under gauging
a discrete $\mathrm{HFS}(q)$. Instead of gauging this symmetry everywhere, one gauges
it only on one side of a codimension-one interface. The resulting interface can be
topological. However, fusing the interface with its orientation reversal does not necessarily
give the identity. Rather, it produces a condensation defect, or a sum over lower-dimensional
topological sectors. Schematically,
\begin{align}
    \mathcal{D}_{\mathrm{half}}
    \otimes
    \overline{\mathcal{D}}_{\mathrm{half}}
    =
    \mathcal{C}_{\mathrm{cond}},
\end{align}
where \(\mathcal{C}_{\mathrm{cond}}\) is a non-trivial condensate \cite{Choi:2022zal}. This
construction generalizes \textit{Kramers--Wannier} duality to higher-dimensional \(QFT\)s
and is particularly natural in gauge theories with higher-form symmetries.

\noindent
A closely related construction is higher gauging. If a \(q\)-form symmetry can be gauged
not in the whole spacetime but on a submanifold of higher codimension, the resulting
insertion is a condensation defect. These condensation defects are generically
non-invertible. Their fusion rules are richer than ordinary group multiplication because
the fusion coefficients may themselves be topological field theories. Thus, in dimensions
higher than two, the correct algebraic object is often not merely a fusion category, but a
higher fusion category whose morphisms encode defects between defects. Equivalently,
the same generalized symmetry data can often be organized in the language of a
Symmetry Topological Field Theory, or \(\mathrm{SymTFT}\), where topological defects encode
symmetry generators, anomalies, fusion rules and possible obstructions to gauging
\cite{Shao:2023gho,Gagliano:2024off,GarciaEtxebarria:2022vzq}.

\noindent
\noindent
The relevance of this framework becomes especially sharp in theories with dynamical gauge fields, Chern--Simons couplings and axionic interactions. In such theories, the naive Noether current associated with a classical symmetry may fail to be gauge invariant, or may fail to define a conserved and quantized charge \cite{Montero:2017yja}. A characteristic example is axion electrodynamics, where the topological coupling modifies Gauss' law. In that case one does not have a charge which is simultaneously conserved, gauge invariant and quantized. Instead, for rational symmetry parameters, the corresponding Page-like operator can be made well-defined by coupling it to an auxiliary \(TQFT\) sector. This produces a topological operator with non-group-like fusion, namely a non-invertible symmetry operator.
The same mechanism appears in theories with \textit{Adler--Bell--Jackiw} anomalies, where a classical chiral symmetry is not simply destroyed, but can survive as a non-invertible topological defect for rational chiral rotations \cite{Cordova:2022ieu, Choi:2022jqy, Choi:2022fgx}.

\noindent
In the case of a brane-source, \textit{Maxwell} and Page notions of charge are distinguished most clearly in \cite{Marolf:2000cb}. In particular, Page charges are conserved, localized and quantized under suitable assumptions, but they are not invariant under large gauge transformations \cite{Kalkkinen:2002tk}. This distinction is central for the construction of non-invertible symmetry operators in
Supergravity \cite{Garcia-Valdecasas:2023mis}, \cite{Fernandez-Melgarejo:2024ffg}. Page-like currents provide conserved and quantized charges, but their dependence on gauge potentials makes the corresponding fractional operators sensitive to large gauge transformations. Consequently, these operators do not define genuine gauge-invariant topological defects by themselves. They become well-defined only after being dressed by an auxiliary topological sector.

\noindent
The rest of this paper is organized as follows. In section
\ref{SECCION: 2 Review Non-Invertible Symmetries of Eleven Dimensional Supergravity}
we review the construction of non-invertible defects in eleven-dimensional Supergravity,
following \cite{Garcia-Valdecasas:2023mis}. In section
\ref{SECCION: 3From Eleven-Dimensional Defects to textit{Kaluza--Klein} Fibrations}
we introduce the \textit{Kaluza--Klein} fibration and study the descent of the
eleven-dimensional electric non-invertible defect to Type IIA Supergravity. In section
\ref{SECCION: 4  The non-invertible sector from the twisted}
we analyze the compactification of the eleven-dimensional \textit{Bianchi} sector, showing
that it splits into an invertible \(H_{[3]}\)-sector and a twisted non-invertible
\(\widetilde F_{[4]}\)-sector. In section
\ref{SECCION: 5 textit{Kaluza--Klein} origin of the fusion rules of the twisted-textit{Bianchi} defect}
we study the resulting categorical structure, including the fusion rules of the compactified
defects and the Picard subgroup generated by the invertible \(H_{[3]}\)-operators. In section
\ref{SECCION: 6 Charged probes and the M-theory/Type IIA dictionary}
we identify the charged probes using the M-theory/Type IIA brane dictionary and compare
with the brane-probe interpretation of \cite{Garcia-Valdecasas:2023mis}. Finally, in section
\ref{SECCION: 7 Discussion and conclusions}
we summarize the main conclusions and discuss the interpretation of the Type IIA defects
as compactified remnants of the eleven-dimensional defect structure.

\section{Review Non-Invertible Symmetries of Eleven Dimensional Supergravity}\label{SECCION: 2 Review Non-Invertible Symmetries of Eleven Dimensional Supergravity}
\noindent
\noindent
The mechanism behind non-invertible symmetry defects appears naturally in
Supergravity, where the gauge sector contains higher-degree differential form potentials
and the corresponding charged objects are extended branes. In Type II theories these
charged objects are \(Dp\)-branes, whereas in eleven-dimensional Supergravity they are
M-branes. In particular, eleven-dimensional Supergravity \cite{Cremmer:1978km}
contains, in its bosonic sector, the metric \(g\) and the three-form potential
\(C_{[3]}\), with field strength
\begin{align}\label{ECUACION: CURVATUA DE ONCE DIMENSIONES}
    F^{(11)}_{[4]} = dC^{(11)}_{[3]} .
\end{align}
The bosonic action is
\begin{align}\label{ECAUCION: ACCIÓN SUPERGRAVEDAD EN ONCE DIMENSIONES}
    S_{11}(g,C^{(11)}_{[3]})
    =
    \frac{1}{2\kappa_{11}^2}
    \int_{\mathcal{M}_{11}}
    \left(
        R\star_{11}\mathbf{1}
        -
        \frac{1}{2}
        F^{(11)}_{[4]}\wedge \star_{11}F^{(11)}_{[4]}
    \right)
    +
    S_{\mathrm{CS}}^{(11)} .
\end{align}
Here \(4\pi \kappa_{11}^2=(2\pi l_p)^9\). The \textit{Chern--Simons} coupling
\begin{align}
    S_{\mathrm{CS}}^{(11)}
    \sim
    \int_{\mathcal{M}_{11}}
   C^{(11)}_{[3]}\wedge F^{(11)}_{[4]}\wedge F^{(11)}_{[4]}
\end{align}
modifies the equation of motion to
\begin{align}
    d\star_{11}F^{(11)}_{[4]}
    +
    \frac{1}{2}
    F^{(11)}_{[4]}\wedge F^{(11)}_{[4]}
    =
    0 .
\end{align}
Therefore, the Maxwell-like current \(\star_{11}F^{(11)}_{[4]}\) is not closed whenever
\(F^{(11)}_{[4]}\wedge F^{(11)}_{[4]}\neq 0\)
\cite{Heidenreich:2020pkc,Montero:2017yja}. The locally conserved object is instead
the \textit{Page} form
\begin{align}
    J_{\mathrm{Page}}^{(7)}
    =
    \star_{11}F^{(11)}_{[4]}
    +
    \frac{1}{2}
    C^{(11)}_{[3]}\wedge F^{(11)}_{[4]},
\end{align}
which satisfies
\begin{align}
    dJ_{\mathrm{Page}}^{(7)}=0
\end{align}
in the absence of explicit sources. However, \(J_{\mathrm{Page}}^{(7)}\) depends on the
gauge potential \(C^{(11)}_{[3]}\), and is therefore sensitive to large gauge transformations.
This is the basic obstruction to interpreting the corresponding fractional symmetry
operator as an ordinary invertible symmetry
\cite{Garcia-Valdecasas:2023mis,Fernandez-Melgarejo:2024ffg}.

\noindent
The construction introduced in \cite{Garcia-Valdecasas:2023mis} shows that this obstruction can be resolved by promoting the would-be fractional higher-form symmetry operator to a non-invertible topological defect. For a rational parameter \(q/N\), the naive fractional operator is schematically
\begin{align}
    U_{q/N}^{e}(\Sigma_7)
    =
    \exp\!\left[
    2\pi i\,\frac{q}{N}
    \int_{\Sigma_7}
    \star F_{[4]}
    \right].
\end{align}
By itself, this object is not a well-defined gauge-invariant topological operator in the full Supergravity theory. The correct defect is obtained by tensoring it with a seven-dimensional auxiliary $TQFT$ coupled to \(F^{(11)}_{[4]}\) as
\begin{align}
    \mathrm{HFS}(3)\xrightarrow{\otimes~TQFT} \widehat{\mathrm{HFS}}(3),
\end{align}
in a local presentation, one writes the defect operator for $q=1$ \cite{Garcia-Valdecasas:2023mis}:
\begin{align}\label{ECUACION: DEFECTO EN ONCE DIMENSIONES}
    \mathcal{D}^{(11)}_{1/N}(\Sigma_7)
    ={}&
    \int \mathcal{D}[\widehat{\Phi}_{[3]}]\,
    \exp\bigg(
    2\pi i
    \int_{\Sigma_7}
    \frac{1}{N}\star_{11} F^{(11)}_{[4]}
    +
    \frac{N}{2}\,
    \widehat{\Phi}_{[3]}\wedge d\widehat{\Phi}_{[3]}
    -
    \widehat{\Phi}_{[3]}\wedge F^{(11)}_{[4]}
    \bigg).
\end{align}
where \(\widehat{\Phi}_{[3]}\) is an auxiliary three-form gauge field living on the defect, and
\(\gcd(q,N)=1\). The precise sign convention in the mixed coupling depends on the convention chosen for the eleven-dimensional \textit{Chern--Simons} term and for the Page current. What is invariant is the mechanism: a fractional Page-like transformation must be dressed by a $TQFT$ in order to define a gauge-invariant topological defect.

\noindent
The fusion of these defects is not group-like. Rather, it is a fusion valued in
topological sectors. When two defects are brought together, the resulting defect contains
the fractional operator with the summed angle together with the auxiliary topological
theory required for gauge invariance. This is the same categorical mechanism that appears
in higher gauging and condensation defects, where fusion coefficients are not ordinary
numbers but lower-dimensional \(TQFT\)s
\cite{Choi:2022zal,Shao:2023gho,Roumpedakis:2022aik}. One has
\begin{align}\label{ECUACION: REGLA FE FUSION DEFECTO EN ONCE DIMENSIONES}
    \mathcal{D}_{q/N}^{(11)}(\Sigma_7)
    \otimes
    \mathcal{D}_{\ell/N}^{(11)}(\Sigma_7)
    \cong
    \mathcal{T}_{BF}^{(N,q-\ell)}(\Sigma_7)
    \otimes
    \mathcal{D}_{(q+\ell)/N}^{(11)}(\Sigma_7),
\end{align}
where \(\mathcal{T}_{BF}^{(N,q-\ell)}(\Sigma_7)\) denotes the residual \(BF\)-type
topological sector supported on the defect worldvolume. Its compactified analogue will be made explicit in section \ref{SECCION: 5 textit{Kaluza--Klein} origin of the fusion rules of the twisted-textit{Bianchi} defect}. For the eleven-dimensional construction, see \cite{Garcia-Valdecasas:2023mis}. Here
\(\gcd(q,N)=\gcd(\ell,N)=1\). Therefore, the fusion is not an ordinary group product. In
particular,
\begin{align}
    \mathcal{D}_{q/N}^{(11)}
    \otimes
    \left(\mathcal{D}_{q/N}^{(11)}\right)^{\dagger}
    \neq
    \mathbf{1}.
\end{align}
Instead, the adjoint fusion produces a non-trivial condensation sector
\(\mathcal{C}\). This is the categorical signature of non-invertibility.

\noindent
The physical action of these Supergravity defects is detected by probe branes. In eleven
dimensions, the M2-brane is electrically charged under \(C_{[3]}\), while the M5-brane is
magnetically charged. The non-invertible defect acts non-trivially on brane operators by
linking. For instance, an M2-brane operator linked with
\(\mathcal{D}^{(11)}_{q/N}(\Sigma_7)\) acquires a phase, or is mapped into a topological
sector determined by its charge and by the coupling to the auxiliary \(TQFT\). Thus, the
non-invertibility is not only an algebraic property of the fusion rule. It is also visible in
the action of the defect on extended charged objects.

\noindent
Type II Supergravities exhibit analogous structures. In the case of SUGRA\((10,2A)\) and
SUGRA\((10,2B)\)\footnote{Here SUGRA\((d,\mathcal N)\) denotes a \(d\)-dimensional Supergravity theory with
supersymmetry type \(\mathcal N\). In ten dimensions, \(2A\) labels the non-chiral Type
IIA theory, with two Majorana--Weyl supersymmetry parameters of opposite chirality,
whereas \(2B\) labels the chiral Type IIB theory, with two Majorana--Weyl parameters
of the same chirality \cite{Becker:2006dvp}.}, the \(NSNS\) and \(RR\) field strengths obey coupled Bianchi
identities and equations of motion. In democratic notation
\cite{Bergshoeff:1996ui}, the \(RR\) field strengths are naturally arranged into
polyforms, and the presence of \(B_{[2]}\) leads to Page-like combinations of the form
\begin{align}
    F_{\mathrm{Page}}
    =
    e^{-B_{[2]}}\wedge F_{\mathrm{RR}} .
\end{align}
The \(B\)-twisted structure of \(RR\) charges is also naturally related to the
K-theoretic classification of D-brane charges
\cite{Minasian:1997mm,Witten:1998cd}. As in eleven dimensions, these Page charges
are localized and quantized, but they are sensitive to large \(B\)-field gauge
transformations \cite{Marolf:2000cb}. Consequently, fractional operators built from
Page-like currents generally require topological dressing. This is the mechanism that
leads to non-invertible defects in ten-dimensional Supergravity.

\noindent
In SUGRA$(10,2B)$, the situation is even richer because the theory possesses an \(\mathrm{SL}(2,\mathbb{Z})\) duality structure acting on the axio-dilaton and on the $NSNS-RR$ two-form doublet. Recent work has constructed non-invertible topological operators associated with the \(0\)-, \(2\)-, \(4\)- and \(6\)-form symmetries of SUGRA$(10,2B)$. The construction proceeds by first identifying \(\mathrm{SL}(2,\mathbb{Z})\)-covariant conserved currents and then introducing fractional, gauge-invariant topological operators by coupling them to appropriate auxiliary $TQFT$s \cite{Fernandez-Melgarejo:2024ffg}. These defects act on the spectrum of BPS branes and can also be formulated in the \textbf{SymTFT} language of half higher-gauging \cite{Brennan:2024fgj}. This Supergravity picture is also tied to the broader quantum-gravity principle that exact global symmetries should not exist in a consistent theory of quantum gravity \cite{Harlow:2018tng}, \cite{Banks:2010zn}. HFS in low-energy effective field theory are often broken or gauged by the presence of dynamical charged branes. In particular, \textit{Chern--Weil} currents such as \cite{Marolf:2000cb}:
\begin{align}
    F_{[2]}\wedge H_{[3]},
    \qquad
    \mathrm{tr}(F_{[2]}\wedge F_{[2]}),
\end{align}
are conserved by \textit{Bianchi} identities and therefore look like robust generalized global symmetries in effective field theory. String theory avoids exact global \textit{Chern--Weil} symmetries through mechanisms such as axions, \textit{Chern--Simons} couplings \cite{Montero:2017yja}, branes ending on branes and branes dissolving into flux. Non-invertible symmetries refine this picture: what appears as an exact fractional symmetry in the low-energy Supergravity description may be removed, gauged or completed by additional UV degrees of freedom in string theory or M-theory \cite{Heckman:2024obe}.

\noindent
For the purposes of the present work, the central lesson is the following. Non-invertible symmetries in Supergravity arise when the naive topological operator associated with a higher-form current fails to be simultaneously conserved, quantized and gauge invariant. The obstruction is usually produced by \textit{Chern--Simons} terms, modified \textit{Bianchi} identities or axionic couplings. The resolution is to dress the fractional operator by an auxiliary $TQFT$. The resulting object is topological, gauge invariant and physically meaningful, but it is no longer invertible under fusion. Therefore, the symmetry algebra of Supergravity is not always a group. It can be a categorical structure whose generators are topological defects and whose fusion rules encode the interplay between flux quantization, Page charges, large gauge transformations and brane charges \cite{Kalkkinen:2002tk}.

\section{From Eleven-Dimensional Defects to \textit{Kaluza--Klein} Fibrations}\label{SECCION: 3From Eleven-Dimensional Defects to textit{Kaluza--Klein} Fibrations}

\noindent
The previous discussion shows that the natural language for non-invertible symmetries in Supergravity is not the language of ordinary global transformations, but the language of topological defects and their fusion rules. In eleven dimensions, this structure is tied to the dynamics of the three-form potential \(C^{(11)}_{[3]}\), to its Page current \footnote{
A Page current is an improved conserved current whose charge is localized and quantized,
but generally not invariant under large gauge transformations because it depends on gauge
potentials.} and to the \textit{Chern--Simons} coupling of the low-energy M-theory action \cite{Cremmer:1978km}. The next question is whether this non-invertible structure is stable under compactification. More precisely, if SUGRA\((11,1)\) is compactified \textit{à la} \textit{Kaluza--Klein} on a circle and becomes Type IIA Supergravity at low energies, one should ask whether the eleven-dimensional non-invertible defect descends to a ten-dimensional defect. This is the point where the \textit{Kaluza--Klein} fibration \cite{Kaste:2002xs},  becomes essential.

\noindent
We therefore consider an eleven-dimensional spacetime \(\mathcal{M}_{11}\) which is the total space of a circle fibration over a ten-dimensional base \(\mathcal{M}_{10}\) \cite{Bergman:2004ne} as in the fibration diagram \ref{FIGURA: FIBRACION COMPACTIFICACION}
\begin{equation}\label{FIGURA: FIBRACION COMPACTIFICACION}
\begin{tikzcd}[column sep=large,row sep=large]
S^1
\arrow[r, hook, "\iota"]
&
\mathcal{M}_{11}
\arrow[d, "\pi"]
\\
&
\mathcal{M}_{10}
\end{tikzcd}
\end{equation}

\noindent
Locally, one may choose a coordinate \(y\) along the circle fiber. Globally, however, \(dy\) is not the correct invariant object if the fibration is non-trivial. The correct object is the \textit{Kaluza--Klein} one-form
\begin{align}\label{ECUACION: CONEXION DE LA FIBRACION KK}
    \eta = dy + A_{[1]},
\end{align}
where \(A_{[1]}\) is the connection one-form of the circle fibration and
\begin{align}\label{ECUACION. CURVATURA DE LA FIBRACION KK}
    F_{[2]} = dA_{[1]}
\end{align}
is its curvature. Geometrically, \(A_{[1]}\) is an Ehresmann connection \cite{MR0152974}. It defines a horizontal distribution in \(T\mathcal{M}_{11}\), thereby allowing the decomposition of differential forms into horizontal \(\Omega^{\perp}\) and vertical \(\Omega^{\parallel}\) components. Physically, \(A_{[1]}\) becomes the $RR$ one-form gauge field of SUGRA$(10,2A)$. The \textit{Kaluza--Klein} reduction is performed by restricting to the zero-mode sector along the circle $\mathcal{L}_{\partial_y}(\cdots)=0$, so that all ten-dimensional fields are independent of the coordinate \(y\) in the \textit{Fourier} expansion. This truncation selects the massless Type IIA fields and discards the tower of massive \textit{Kaluza--Klein} modes \cite{Duff:1996aw}. The eleven-dimensional metric is then written in the standard form
\begin{align}
    ds^2_{11}=e^{-2\phi/3}ds^2_{10}+e^{4\phi/3}\eta^2,
\end{align}
where \(ds^2_{10}\) is the ten-dimensional \textit{string-frame} metric and \(\phi\) is the Type IIA dilaton. This ansatz makes explicit that the ten-dimensional dilaton measures the radius of the M-theory circle \cite{Witten:1995ex} \cite{Becker:2006dvp}, \cite{Horava:1995qa}, while the connection of the circle fibration becomes the RR one-form potential. The same horizontal--vertical splitting must be applied to the eleven-dimensional three-form potential. The natural decomposition is
\begin{align}\label{ECUACION: COMPACTIFICACION DEL C3 DE ONCE DIMENSIONES}
    C^{(11)}_{[3]}
    =
    C_{[3]} - B_{[2]}\wedge \eta ,
\end{align}
where \(C_{[3]}\) is the Type IIA $RR$ three-form and \(B_{[2]}\) is the $NSNS$ two-form. This formula is more than a local rewriting. It expresses the fact that the single M-theory gauge field \(C^{(11)}_{[3]}\) contains, after compactification, both the $RR$ and $NSNS$ two sectors of Type IIA. Taking the exterior derivative gives
\begin{align}\label{ECUACION: DESCOMPOSICION CURVATURA DE SUGRA EN ONCE DIMENSIONES}
    F^{(11)}_{[4]}=\widetilde{F}_{[4]}-H_{[3]}\wedge \eta ,
\end{align}
with
\begin{align}\label{ECUACION: ECUACION DE MOVIMIENTO SECTOR RR Y NSNS DIRECTO}
    H_{[3]} &= dB_{[2]},
    \\
    \widetilde F_{[4]} &= dC_{[3]}-B_{[2]}\wedge F_{[2]}.
\end{align}
The eleven-dimensional four-form splits into a horizontal \(RR\) field strength and a vertical component controlled by the \(NSNS\) flux. The \textit{Bianchi} identity (namely the closure condition for the $4$-form field strength) then reduces to the coupled Type IIA \textit{Bianchi} identities. Since
\begin{align}\label{ECUACION: BIANCHI DE ONCE DIMENSIONES}
    dF^{(11)}_{[4]}=0,
\end{align}
one obtains
\begin{align}\label{ECUACION: ECUACIONES DE BIANCHI EN DIEZ DIMENSIONES}
    dH_{[3]} &= 0,\\
    d\widetilde{F}_{[4]} + H_{[3]}\wedge F_{[2]} &=0,
\end{align}
This is the first indication that the HFS structure of Type IIA is not simply inherited as a direct product of independent symmetries whenever the compactification is not restricted to be trivial. The $RR$ and $NSNS$ sectors are mixed by the \textit{Kaluza--Klein} curvature \(F_{[2]}\), and this mixing is inherited directly from the geometry of the \(S^1\)-fibration.
At the level of currents, the same phenomenon occurs. The eleven-dimensional Page current descends into Type IIA Page-like combinations involving \((\widetilde{F}_{[4]}\), \(H_{[3]}\), \(B_{[2]})\) and the $RR$ fields. These ten-dimensional currents are the natural objects for charge quantization and for the coupling to D-brane and fundamental string probes. However, as in eleven dimensions, they are sensitive to large gauge transformations \cite{Kalkkinen:2002tk}. Thus, the obstruction responsible for non-invertibility in eleven dimensions does not disappear after compactification. It is reorganized into the coupled $RR$ and $NSNS$ gauge structure of Type IIA.

\noindent
Let \(\Sigma_7\subset M_{11}\) be the support of the eleven-dimensional non-invertible
defect. To obtain a six-dimensional defect in the ten-dimensional theory, we assume that
\(\Sigma_7\) is compatible with the same circle fibration. Namely, we take
\begin{equation}\label{FIGURA: FIBRACION COMPACTIFICACION SOBRE SOPORTE COMPACTO}
\begin{tikzcd}[column sep=large,row sep=large]
S^1
\arrow[r, hook,]
&
\Sigma_7
\arrow[d, "\pi"]
\\
&
\Sigma_6 .
\end{tikzcd}
\end{equation}
The pushforward along the fiber selects the terms in the defect action with one vertical
factor \(\eta\). Therefore, the auxiliary field must also be decomposed with respect to the same Ehresmann connection. If \(\widehat{\Phi}_{[3]}\) denotes the auxiliary three-form field on \(\Sigma_7\), then
\begin{align}
    \widehat{\Phi}_{[3]}
    =
    \widehat{\varphi}_{[3]}
    +
    \widehat{\varphi}_{[2]}\wedge\eta ,
\end{align}
where \(\widehat{\varphi}_{[3]}\) and \(\widehat{\varphi}_{[2]}\) are horizontal fields on \(\Sigma_6\). The component \(\widehat{\varphi}_{[3]}\) remains a three-form on the base, while the component with one leg along the fiber becomes a two-form on the base.

\noindent
Substituting this decomposition into the seven-dimensional auxiliary topological action \eqref{ECUACION: DEFECTO EN ONCE DIMENSIONES}, one keeps only the terms with exactly one vertical factor \(\eta\). Since the vertical one-form \(\eta\) and its curvature \(F_{[2]}\) were already introduced in \eqref{ECUACION: CONEXION DE LA FIBRACION KK} and \eqref{ECUACION. CURVATURA DE LA FIBRACION KK}, respectively, the \textit{pushforward} of the auxiliary sector depends on whether the circle fibration is topologically trivial or not. In the globally non-trivial case. \(F_{[2]}\) represents the \textit{de Rham} curvature of the principal \(U(1)\)-bundle given by the diagram \ref{FIGURA: FIBRACION COMPACTIFICACION SOBRE SOPORTE COMPACTO}, and its integral cohomology class is identified with the first \textit{Chern} class,
\begin{align}
    [\frac{1}{2\pi}F_{[2]}]
    =
    c_1(\Sigma_7)
    \in
    H^2(\Sigma_6,\mathbb{Z}) .
\end{align}
Thus, keeping \(F_{[2]}\neq 0\) amounts to retaining the effect of the non-trivial \textit{Kaluza--Klein} fibration on the auxiliary defect theory. With this understood, the \textit{Chern--Simons}-like term of the auxiliary sector reduces, up to total derivatives on \(\Sigma_6\), as
\begin{align}
    \frac{N}{2}
    \int_{\Sigma_7}
    \widehat{\Phi}_{[3]}\wedge d\widehat{\Phi}_{[3]}
    \quad
    \longrightarrow
    \quad
    N
    \int_{\Sigma_6}
    \widehat{\varphi}_{[3]}\wedge d\widehat{\varphi}_{[2]}
    +
    \frac{N}{2}
    \int_{\Sigma_6}
    \widehat{\varphi}_{[2]}\wedge
    \widehat{\varphi}_{[2]}\wedge F_{[2]} .
\end{align}
The first term is the \(BF\)-type coupling on the base. The second term is the curvature-dependent deformation induced by the non-triviality of the circle bundle. Hence, when the fibration is treated as locally trivial, or when the curvature-dependent sector is truncated, the auxiliary theory reduces to the pure \(BF\)-type term
\begin{align}
    \frac{N}{2}
    \int_{\Sigma_7}
    \widehat{\Phi}_{[3]}\wedge d\widehat{\Phi}_{[3]}
    \quad
    \longrightarrow
    \quad
    N
    \int_{\Sigma_6}
    \widehat{\varphi}_{[3]}\wedge d\widehat{\varphi}_{[2]} .
\end{align}
For a genuinely non-trivial \(S^1\)-fibration, however, the compactified auxiliary sector is a \(BF\)-type topological theory deformed by the first Chern class of the Kaluza--Klein circle bundle. This is the higher-form version of the familiar mechanism by which \textit{Chern--Simons} theories on circle bundles reduce to \(BF\)-type topological theories on the base, with additional curvature-dependent terms when the bundle is non-trivial \cite{Mickler:2015eca,Marolf:2000cb,Ishii:2007sy}. In the present case, the mechanism acts on the auxiliary \(TQFT\) sector of the non-invertible defect.

\noindent
The mixed coupling between the auxiliary field and the eleven-dimensional four-form
descends in the same way. With the sign convention adopted in the eleven-dimensional
defect, this term is
\begin{align}
    -
    \int_{\Sigma_7}
    \widehat{\Phi}_{[3]}\wedge F^{(11)}_{[4]} .
\end{align}
Using the decompositions of the auxiliary field and of the four-form field strength, the
terms with one vertical factor \(\eta\) give
\begin{align}
    -
    \int_{\Sigma_7}
    \widehat{\Phi}_{[3]}\wedge F^{(11)}_{[4]}
    \quad
    \longrightarrow
    \quad
    -
    \int_{\Sigma_6}
    \widehat{\varphi}_{[2]}\wedge \widetilde{F}_{[4]}
    +
    \int_{\Sigma_6}
    \widehat{\varphi}_{[3]}\wedge H_{[3]} .
\end{align}
\noindent
Thus, the \(RR\) field strength \(\widetilde{F}_{[4]}\) couples to the two-form
component of the auxiliary field, while the \(NSNS\) field strength \(H_{[3]}\) couples
to its three-form component. This is the structure naturally produced by the
Kaluza--Klein split: the auxiliary sector restores the gauge consistency of the
fractional \(RR\) insertion and keeps track of the \(NSNS\) mixing inherited from
eleven dimensions.

\noindent
In the electric defect sector below, we keep the pure \(BF\)-type part of the compactified auxiliary theory and neglect the additional curvature-dependent self-coupling of the auxiliary field. This does not mean that \(F_{[2]}\) is set to zero in the bulk Kaluza--Klein reduction. The curvature \(F_{[2]}\) is still retained in the Type IIA \textit{Bianchi} identities, where it is responsible for the twisted identity
\begin{align}
    d\widetilde F_{[4]}+H_{[3]}\wedge F_{[2]}=0 .
\end{align}
Thus, the truncation concerns only the curvature deformation of the auxiliary electric
defect sector, not the full \textit{Kaluza--Klein} background. In this sector the auxiliary \textit{Chern--Simons}-like term
reduces to a pure \(BF\)-type coupling on \(\Sigma_6\). In this sector, the eleven-dimensional defect descends to the effective Type IIA defect \cite{Garcia-Valdecasas:2023mis}
\begin{align}
    \mathcal{D}^{(10),e}_{p/N}(\Sigma_6)
    ={}&
    \int
    \mathcal{D}[\widehat{\varphi}_{[2]}]\,
    \mathcal{D}[\widehat{\varphi}_{[3]}]\,
    \exp\bigg(
    2\pi i
    \int_{\Sigma_6}
    \frac{p}{N}\star_{10}\widetilde{F}_{[4]}
    +
    N\,\widehat{\varphi}_{[3]}\wedge d\widehat{\varphi}_{[2]}
    \notag
    \\
    &\hspace{2.1cm}
    -
    \widehat{\varphi}_{[2]}\wedge\widetilde{F}_{[4]}
    +
    p\,\widehat{\varphi}_{[3]}\wedge H_{[3]}
    \bigg),
    \notag
    \\
    ={}&
    U_{p/N}(\Sigma_6)
    \otimes
    \mathcal{A}_{TQFT}^{(N,p)}
    \!\left(
        \frac{\widetilde{F}_{[4]}}{N}
    \right).
\end{align}
Here \(p/N\) is the fractional parameter inherited from the eleven-dimensional operator,
with \(\gcd(p,N)=1\). The first term is the fractional \(RR\) electric insertion. In writing it as
\(\star_{10}\widetilde F_{[4]}\), we suppress the dilaton-dependent normalization factors arising from the explicit reduction of the eleven-dimensional Hodge star. These factors are irrelevant for the topological defect algebra discussed here, and can be restored from the standard Type IIA \textit{Kaluza--Klein} reduction of the eleven-dimensional metric. The second term is the compactified \(BF\)-type auxiliary sector. The last two terms are the mixed couplings to the \(RR\) and \(NSNS\) field strengths, with their relative signs fixed by the convention chosen for the \textit{Kaluza--Klein} decomposition of \(C^{(11)}_{[3]}\) and \(F^{(11)}_{[4]}\).

\noindent
If the \(U(1)\)-fibration \ref{FIGURA: FIBRACION COMPACTIFICACION SOBRE SOPORTE COMPACTO} is kept globally non-trivial, then \(F_{[2]}\) represents the curvature
of the \textit{Kaluza--Klein} circle bundle and the auxiliary sector receives an additional
curvature dependent topological correction proportional to
\begin{align}
    \int_{\Sigma_6}
    \widehat{\varphi}_{[2]}\wedge
    \widehat{\varphi}_{[2]}\wedge F_{[2]} .
\end{align}
This term deforms the \(BF\)-type auxiliary theory, but it will not be included in the
defect sector analyzed below. Thus, the results that follow are understood in the
locally trivial or curvature-truncated Kaluza--Klein sector.
\noindent
The construction can be summarized by the commutative diagram
\begin{equation}
\begin{tikzcd}[column sep=large,row sep=large]
\mathrm{SUGRA}(11,1)
\arrow[r, "\mathrm{KK}"]
\arrow[d, "\mathcal{D}^{(11)}_{q/N}"']
&
\mathrm{SUGRA}(10,2A)
\arrow[d, "\mathcal{D}^{(10)}_{p/N}"]
\\
\mathrm{Def}^{\mathrm{non\text{-}inv}}_{11d}
\arrow[r, "\pi_*"']
&
\mathrm{Def}^{\mathrm{non\text{-}inv}}_{10d}
\end{tikzcd}
\end{equation}
where the upper horizontal arrow is the usual \textit{Kaluza--Klein} fibration of the bulk theory, while the lower horizontal arrow is the induced \textit{pushforward} of the non-invertible defect. The diagram expresses the main physical statement of this paper: compactification
transports the categorical defect structure from eleven dimensions to Type IIA in ten
dimensions. The non-invertibility of the ten-dimensional defect is inherited from the eleven-dimensional \textit{Chern--Simons} obstruction, not generated accidentally in the lower-dimensional theory.

\noindent
The existence of the eleven-dimensional non-invertible defect follows from the known
construction of non-invertible symmetries in Supergravity \cite{Garcia-Valdecasas:2023mis}. In the present work, we study its \textit{Kaluza--Klein} reduction by treating the defect support as an \(S^1\)-fibration and by
decomposing the auxiliary topological field together with the bulk fields. Under this
reduction, the Type IIA non-invertible defect is obtained as the \textit{pushforward} of the
eleven-dimensional defect along the circle fiber. In this sense, it can be interpreted as the
compactified counterpart of the eleven-dimensional defect structure.

\section{Compactification of the invertible Bianchi sector}\label{SECCION: 4  The non-invertible sector from the twisted}

\noindent
We now turn to the sector controlled by the eleven-dimensional \textit{Bianchi} identity,
\begin{align}\label{ECUACION: ECUACION DE BIANCHI EN ONCE DIMENSIONES}
    dF^{(11)}_{[4]}=0.
\end{align}
In eleven dimensions, this identity defines an ordinary invertible magnetic HFS$(6)$. Indeed, for a closed four-cycle \(\mathcal{Q}_4\subset M_{11}\), one can introduce
the magnetic topological operator
\begin{align}
    U^{(11),m}_{\beta}(\mathcal{Q}_4)
    =
    \exp\!\bigg(
    2\pi i\,\beta
    \int_{\mathcal{Q}_4}
    F^{(11)}_{[4]}
    \bigg),
    \qquad
    \beta\in \mathbb{R}/\mathbb{Z}.
\end{align}
Since \(F^{(11)}_{[4]}\) is closed, this operator is invariant under smooth deformations
of \(\mathcal{Q}_4\), as long as no magnetic source is crossed. Moreover, its fusion is
group-like:
\begin{align}
    U^{(11),m}_{\beta}
    \otimes
    U^{(11),m}_{\beta'}
    =
    U^{(11),m}_{\beta+\beta'},
    \qquad
    \left(
    U^{(11),m}_{\beta}
    \right)^{-1}
    =
    U^{(11),m}_{-\beta}, \qquad \beta, \beta' \in \mathbb{R}/\mathbb{Z}.
\end{align}
Therefore, in the source-free Supergravity regime, the magnetic \textit{Bianchi} sector realizes an invertible magnetic HFS with group $U(1)^{(6)}_{m}.$ The degree \(6\) refers to the dimension of the charged operator, namely the six-dimensional worldvolume of the M5-brane. Equivalently, the support of the symmetry operator has dimension \(4\), as required by $4+6=10=11-1$. We now compactify this sector using the \textit{Kaluza--Klein} decomposition of the
four-form field strength of SUGRA\((11,1)\) introduced in
\eqref{ECUACION: DESCOMPOSICION CURVATURA DE SUGRA EN ONCE DIMENSIONES}.
The single eleven-dimensional closure condition
\eqref{ECUACION: ECUACION DE BIANCHI EN ONCE DIMENSIONES}
does not remain a single identity in Type IIA. Instead, it decomposes into the two
\textit{Bianchi} identities displayed in
\eqref{ECUACION: ECUACIONES DE BIANCHI EN DIEZ DIMENSIONES}. This splitting is the
key point: the eleven-dimensional invertible \textit{Bianchi} sector does not descend to one
isolated ten-dimensional invertible symmetry. Rather, it decomposes into two different
sectors.

\noindent
The first sector is the vertical component of \eqref{ECUACION: ECUACIONES DE BIANCHI EN DIEZ DIMENSIONES}, governed by \(dH_{[3]}=0\), and remains an
ordinary invertible \(NSNS\) magnetic sector. The second sector is the horizontal \(RR\)
component, governed by the twisted identity
\begin{align}
    d\widetilde F_{[4]} + H_{[3]}\wedge F_{[2]}=0.
\end{align}
Because \(\widetilde F_{[4]}\) is not closed by itself whenever
\(H_{[3]}\wedge F_{[2]}\neq 0\), a bare operator built only from
\(\widetilde F_{[4]}\) is not a genuine invertible topological operator in a generic
\textit{Kaluza--Klein} fibration \cite{Apruzzi:2023uma}. Thus, after compactification, the originally invertible eleven-dimensional magnetic sector splits into an invertible \(H_{[3]}\)-sector and
a twisted \(RR\) sector that requires a topological dressing by an auxiliary
\(BF\)-type sector \cite{Garcia-Valdecasas:2023mis}, rather than a completion of the physical charge spectrum.

\noindent
The component of \(F^{(11)}_{[4]}\) \eqref{ECUACION: CURVATUA DE ONCE DIMENSIONES} with one leg along the circle gives the $NSNS$ field strength \(H_{[3]}\). If the eleven-dimensional four-cycle wraps the M-theory circle,
\begin{align}
    \mathcal{Q}_4=S^1\times \mathcal{Q}_3,
\end{align}
then
\begin{align}
    \int_{\mathcal{Q}_4}
    F^{(11)}_{[4]}
    ={}&
    -
    \int_{S^1\times\mathcal{Q}_3}
    H_{[3]}\wedge\eta,\notag
    \\
    ={}&
    -
    \int_{\mathcal{Q}_3}
    H_{[3]},
\end{align}
up to the normalization and orientation of the circle fiber. Therefore, after absorbing the
orientation sign into the parameter \(\beta\), the wrapped part of the eleven-dimensional
magnetic operator descends to the Type IIA invertible operator
\begin{align}
    U^{(10),H}_{\beta}(\mathcal{Q}_3)= \exp\bigg(2\pi i\,\beta\int_{\mathcal{Q}_3}H_{[3]}\bigg).
\end{align}
Since
\begin{align}
    dH_{[3]}=0,
\end{align}
this operator is topological and invertible:
\begin{align}
    U^{(10),H}_{\beta}
    \otimes
    U^{(10),H}_{\beta'}
    =
    U^{(10),H}_{\beta+\beta'},
    \qquad
    \left(
    U^{(10),H}_{\beta}
    \right)^{-1}
    =
    U^{(10),H}_{-\beta}, \quad \beta, \beta' \in \mathbb{R}/\mathbb{Z}.
\end{align}
This is the first sector produced by compactifying the eleven-dimensional invertible
\textit{Bianchi} operator. It remains an ordinary invertible magnetic HFS$(6)$ operator in Type IIA. In the source-free regime, it generates the magnetic
\(B_{[2]}\)-sector
\begin{align}
    U(1)^{(6),B_{[2]}}_{m}.
\end{align}
The degree \(6\) is fixed by the dimension of the charged object, namely the
six-dimensional worldvolume of the \(NS5\)-brane.

\subsection{The non-invertible sector from the twisted \textit{Bianchi} identity}

\noindent
The horizontal component of the eleven-dimensional \textit{Bianchi} identity gives the
second compactified sector. Using the \textit{Kaluza--Klein} decomposition of \(F^{(11)}_{[4]}\) in \eqref{ECUACION: DESCOMPOSICION CURVATURA DE SUGRA EN ONCE DIMENSIONES}, together with the Type IIA \textit{Bianchi} identities in \eqref{ECUACION: ECUACIONES DE BIANCHI EN DIEZ DIMENSIONES}, one sees that the horizontal \(RR\) field strength \(\widetilde F_{[4]}\) is not closed by itself. A naive horizontal reduction of the eleven-dimensional magnetic operator would therefore suggest the insertion
\begin{align}
    U^{(10),\widetilde F}_{\gamma}(\mathcal{Q}_4)
    =
    \exp\!\left[
    2\pi i\,\gamma
    \int_{\mathcal{Q}_4}
    \widetilde F_{[4]}
    \right],
    \qquad
    \gamma\in \mathbb{R}/\mathbb{Z}.
\end{align}
However, due to the twisted identity
\begin{align}
    d\widetilde F_{[4]}
    =
    -
    H_{[3]}\wedge F_{[2]},
\end{align}
this operator is not topological in a generic \textit{Kaluza--Klein} fibration. Indeed,
if \(\mathcal{Q}_4\) is deformed through a five-chain \(\mathcal{V}_5\), the exponent changes by
\begin{align}
    \int_{\mathcal{V}_5}
    d\widetilde F_{[4]}
    =
    -
    \int_{\mathcal{V}_5}
    H_{[3]}\wedge F_{[2]} .
\end{align}
Thus, the bare horizontal operator is a genuine invertible topological operator only when
the twisting class \(H_{[3]}\wedge F_{[2]}\) vanishes. Otherwise, it must be replaced by a
topologically dressed operator, obtained by coupling the bare insertion to an auxiliary
\(BF\)-type sector. This is precisely the Type IIA Supergravity mechanism for
non-invertible defects analyzed in \cite{Garcia-Valdecasas:2023mis}. In the present
construction, we show that the same dressed defect is obtained as the horizontal component of the
compactified eleven-dimensional magnetic sector.

\noindent
The corresponding fractional defect is
\begin{align}
    \mathcal{D}^{(10),\widetilde F}_{r/N}(\mathcal{Q}_4)
    ={}&
    \int
    \mathcal{D}[\widehat{\chi}_{[1]}]\,
    \mathcal{D}[\widehat{\chi}_{[2]}]\,
    \exp\bigg(
    2\pi i
    \int_{\mathcal{Q}_4}
        \frac{r}{N}\,\widetilde F_{[4]}
        +
        N\,\widehat{\chi}_{[2]}\wedge d\widehat{\chi}_{[1]}
        +\notag
        \\
        +{}&
        \widehat{\chi}_{[2]}\wedge F_{[2]}
        +
        r\,\widehat{\chi}_{[1]}\wedge H_{[3]}
\bigg)
\end{align}
With $\gcd(r,N)=1 $, here \(\widehat{\chi}_{[1]}\) and \(\widehat{\chi}_{[2]}\) are auxiliary fields supported on the four-dimensional defect worldvolume \(\mathcal{Q}_4\). The term
\begin{align}
    N\int_{\mathcal{Q}_4}\widehat{\chi}_{[2]}\wedge d\widehat{\chi}_{[1]}
\end{align}
is the \(BF\)-type auxiliary topological theory, while the mixed couplings
\begin{align}
    \int_{\mathcal{Q}_4}
    \widehat{\chi}_{[2]}\wedge F_{[2]},
    \qquad
    r
    \int_{\mathcal{Q}_4}
    \widehat{\chi}_{[1]}\wedge H_{[3]}
\end{align}
encode the \(RR-NSNS\) mixing responsible for the non-closure of \(\widetilde F_{[4]}\). Therefore, the twisted horizontal component of the compactified \textit{Bianchi} identity gives a non-invertible Type IIA defect. In this formulation, the defect is not introduced independently in ten dimensions: it is the compactified horizontal remnant of the eleven-dimensional magnetic \textit{Bianchi} sector.

\noindent
This result is important because it shows that even the eleven-dimensional sector that was invertible before compactification can produce a non-invertible sector in ten dimensions. The reason is purely geometric: the horizontal $RR$ field strength \(\widetilde F_{[4]}\) is not closed independently. Its \textit{Bianchi} identity is twisted by the product \(H_{[3]}\wedge F_{[2]}\). Hence, the \textit{Kaluza--Klein} fibration splits the eleven-dimensional invertible magnetic defect into
\begin{align}
    U^{(11),m}_{\beta}(\mathcal{Q}_4)
    \quad
    \xrightarrow{\;\mathrm{KK}\;}
    \quad
    \left(
        U^{(10),H}_{\beta}(\mathcal{Q}_3),
        \,
        U^{(10),\widetilde F}_{\beta}(\mathcal{Q}_4)
    \right).
\end{align}
For rational values \(\beta=r/N\), the bare horizontal operator is replaced by the
dressed defect
\begin{align}
    U^{(10),\widetilde F}_{r/N}(\mathcal{Q}_4)
    \quad
    \leadsto
    \quad
    \mathcal{D}^{(10),\widetilde F}_{r/N}(\mathcal{Q}_4).
\end{align}
The first component is invertible because it is controlled by \(dH_{[3]}=0\). The second component is non-invertible because it is controlled by the twisted identity
\begin{align}
    d\widetilde F_{[4]} + H_{[3]}\wedge F_{[2]}=0 .
\end{align}

\noindent
Combining the equation of motion sector and the compactified \textit{Bianchi} sector analyzed here, the relevant set of defects obtained in the ten-dimensional theory are
\begin{align}
    \left\{
    \mathcal{D}^{(10),e}_{p/N}(\Sigma_6),
    \;
    U^{(10),H}_{\beta}(\mathcal{Q}_3),
    \;
    \mathcal{D}^{(10),\widetilde F}_{r/N}(\mathcal{Q}_4)
    \right\}.
\end{align}
Their origins are
\begin{align}
    d\star_{11}F^{(11)}_{[4]}
    +
    \frac{1}{2}
    F^{(11)}_{[4]}\wedge F^{(11)}_{[4]}
    =
    0
    \quad
    &\Longrightarrow
    \quad
    \mathcal{D}^{(10),e}_{p/N}(\Sigma_6),
\end{align}
and
\begin{align}
    dF^{(11)}_{[4]}=0
    \quad
    &\Longrightarrow
    \quad
    \left\{
        U^{(10),H}_{\beta}(\mathcal{Q}_3),
        \;
        \mathcal{D}^{(10),\widetilde F}_{r/N}(\mathcal{Q}_4)
    \right\}.
\end{align}
The first defect is non-invertible because it descends from the eleven-dimensional Page-current obstruction. The second is invertible because it is controlled by the closed $NSNS$ field strength \(H_{[3]}\). The third is non-invertible because the RR field strength \(\widetilde F_{[4]}\) obeys a twisted \textit{Bianchi} identity rather than an ordinary closure condition. Equivalently, the \textit{Kaluza--Klein} compactification produces the following result:
\begin{align*}
    \boxed{
    \begin{array}{ccl}
    \text{11d electric equation of motion}
    &\longrightarrow&
    \text{10d electric non-invertible defect},
    \\[4pt]
    \text{11d Bianchi identity, vertical component}
    &\longrightarrow&
    \text{10d invertible }H_{[3]}\text{ defect},
    \\[4pt]
    \text{11d Bianchi identity, horizontal component}
    &\longrightarrow&
    \text{10d non-invertible }\widetilde F_{[4]}\text{ defect}.
    \end{array}
    }
\end{align*}
This is the refined result of the compactification. It is not only the eleven-dimensional
non-invertible electric defect that survives the reduction. The compactification also reveals that
the apparently invertible eleven-dimensional \textit{Bianchi} sector splits into an invertible $NSNS$ sector
and a non-invertible $RR$ sector. Therefore, the complete ten-dimensional categorical structure is
larger than the direct reduction of the electric Page defect alone. In particular, the \textit{Kaluza--Klein} fibration does not merely reduce fields. It decomposes the defect algebra. The eleven-dimensional data descend to a ten-dimensional system in which invertible and non-invertible sectors coexist:
\begin{align}
    \left(
    dH_{[3]},
    \;
    d\widetilde F_{[4]}+H_{[3]}\wedge F_{[2]},
    \;
    \mathcal{D}^{(10),e}_{p/N}
    \right).
\end{align}
This is the precise sense in which the compactification of the full defect structure produces
both ordinary invertible higher-form operators and genuinely non-invertible topological defects
in SUGRA$(10,2A)$.

\section{\textit{Kaluza--Klein} origin of the fusion rules of the twisted-\textit{Bianchi} defect}\label{SECCION: 5 textit{Kaluza--Klein} origin of the fusion rules of the twisted-textit{Bianchi} defect}

\noindent
The fusion rules shown above should not be understood as a new abstract mechanism for
non-invertible fusion. The appearance of residual topological sectors in the fusion of
non-invertible defects is a standard feature of condensation defects and higher gauging.
The new point here is instead the \textit{Kaluza--Klein} origin of this fusion algebra. The
\(BF\)-dressed defect associated with the twisted Type IIA \textit{Bianchi} identity is obtained
as the horizontal compactified remnant of the eleven-dimensional magnetic \textit{Bianchi}
operator. Consequently, the residual \(BF\)-type sector appearing in the fusion rules is
not introduced independently in ten dimensions, but is inherited from the compactified
auxiliary topological sector of the eleven-dimensional construction.

\noindent
The non-invertible sector associated with the compactified \textit{Bianchi} identity must also be
equipped with its fusion rules. This is important because the equation
\begin{align}
    d\widetilde F_{[4]}+H_{[3]}\wedge F_{[2]}=0
\end{align}
shows that the horizontal \(RR\) field strength \(\widetilde F_{[4]}\) is not closed by itself. Therefore,
the corresponding fractional operator cannot be an ordinary invertible topological operator.
The correct object is the \(TQFT\)-dressed defect
\begin{align}
    \mathcal{D}^{(10),\widetilde F}_{r/N}(\mathcal{Q}_4)
    =
    U^{(10),\widetilde F}_{r/N}(\mathcal{Q}_4)
    \otimes
    \mathcal{A}^{(N,r)}_{\mathrm{BF}}
    \!\left(
        \frac{\widetilde F_{[4]}}{N},
        H_{[3]},
        F_{[2]}
    \right),
\end{align}
where \(U^{(10),\widetilde F}_{r/N}\) denotes the would-be fractional operator
\begin{align}
    U^{(10),\widetilde F}_{r/N}(\mathcal{Q}_4)
    =
    \exp\!\left[
        2\pi i\,\frac{r}{N}
        \int_{\mathcal{Q}_4}
        \widetilde F_{[4]}
    \right],
\end{align}
and \(\mathcal{A}^{(N,r)}_{\mathrm{BF}}\) is the auxiliary \(BF\)-type sector obtained by compactifying
the eleven-dimensional auxiliary \(TQFT\). In a local presentation, this sector is represented by
the auxiliary fields \(\widehat{\chi}_{[1]}\) and \(\widehat{\chi}_{[2]}\) through
\begin{align}
    \mathcal{A}^{(N,r)}_{\mathrm{BF}}
    =
    \int
    \mathcal{D}[\widehat{\chi}_{[1]}]\,
    \mathcal{D}[\widehat{\chi}_{[2]}]\,
    \exp\!\left[
        2\pi i
        \int_{\mathcal{Q}_4}
        \left(
            N\,\widehat{\chi}_{[2]}\wedge d\widehat{\chi}_{[1]}
            +
            \widehat{\chi}_{[2]}\wedge F_{[2]}
            +
            r\,\widehat{\chi}_{[1]}\wedge H_{[3]}
        \right)
    \right].
\end{align}
Thus the complete defect is not determined only by the phase built from \(\widetilde F_{[4]}\). Its definition also contains the \(BF\)-type topological theory which compensates the failure of \(d\widetilde F_{[4]}\) to vanish.

\noindent
The fusion rule follows the same logic as the eleven-dimensional defect. For two rational
defects with \(\gcd(r,N)=\gcd(s,N)=1\), one obtains
\begin{align}
    \mathcal{D}^{(10),\widetilde F}_{r/N}(\mathcal{Q}_4)
    \otimes
    \mathcal{D}^{(10),\widetilde F}_{s/N}(\mathcal{Q}_4),\notag
    ={}&
    \left[
        U^{(10),\widetilde F}_{r/N}(\mathcal{Q}_4)
        \otimes
        \mathcal{A}^{(N,r)}_{\mathrm{BF}}
    \right]
    \otimes
    \left[
        U^{(10),\widetilde F}_{s/N}(\mathcal{Q}_4)
        \otimes
        \mathcal{A}^{(N,s)}_{\mathrm{BF}}
    \right],\notag
    \\
    \simeq{}&
    U^{(10),\widetilde F}_{r/N}(\mathcal{Q}_4)
    \cdot
    U^{(10),\widetilde F}_{s/N}(\mathcal{Q}_4)
    \otimes
    \left[
        \mathcal{A}^{(N,r)}_{\mathrm{BF}}
        \otimes
        \mathcal{A}^{(N,s)}_{\mathrm{BF}}
    \right],\notag
    \\
    ={}&
    U^{(10),\widetilde F}_{(r+s)/N}(\mathcal{Q}_4)
    \otimes
    \left[
        \mathcal{A}^{(N,r)}_{\mathrm{BF}}
        \otimes
        \mathcal{A}^{(N,s)}_{\mathrm{BF}}
    \right].
\end{align}
The important point of this fusion rule is that the last factor does not disappear. The fusion is therefore not the
ordinary group law of the fractional phases. Instead, it is modified by the auxiliary topological
sector inherited from the \textit{Kaluza--Klein} reduction.

\noindent
When the product of \(BF\)-type sectors can be reorganized into a single effective auxiliary sector,
one obtains the compact expression $\mathcal{A}^{(N,r)}_{\mathrm{BF}}\otimes\mathcal{A}^{(N,s)}_{\mathrm{BF}}\leadsto\mathcal{A}^{(N,r+s)}_{\mathrm{BF}}$. In that case the effective fusion rule becomes
\begin{align}
    \mathcal{D}^{(10),\widetilde F}_{r/N}(\mathcal{Q}_4)
    \otimes
    \mathcal{D}^{(10),\widetilde F}_{s/N}(\mathcal{Q}_4)
    \leadsto
    U^{(10),\widetilde F}_{(r+s)/N}(\mathcal{Q}_4)
    \otimes
    \mathcal{A}^{(N,r+s)}_{\mathrm{BF}}.
\end{align}

\noindent
Equivalently, using the definition of the dressed defect, one may write
\begin{align}
    \mathcal{D}^{(10),\widetilde F}_{r/N}(\mathcal{Q}_4)
    \otimes
    \mathcal{D}^{(10),\widetilde F}_{s/N}(\mathcal{Q}_4)
    \overset{\mathrm{eff}}{\cong}
    \mathcal{D}^{(10),\widetilde F}_{[r+s]_{N}/N}(\mathcal{Q}_4),
\end{align}
only after suppressing the residual \(BF\)-type topological sector into the
definition of the dressed defect. Here \([r+s]_{N}\) denotes reduction modulo \(N\).
More explicitly, since
\begin{align}
    \mathcal{D}^{(10),\widetilde F}_{r/N}
    =
    U^{(10),\widetilde F}_{r/N}
    \otimes
    \mathcal{A}^{(N,r)}_{\mathrm{BF}},
\end{align}
one finds
\begin{align}
    \mathcal{D}^{(10),\widetilde F}_{r/N}(\mathcal{Q}_4)
    \otimes
    \mathcal{D}^{(10),\widetilde F}_{s/N}(\mathcal{Q}_4)
    \cong{}&
    U^{(10),\widetilde F}_{[r+s]_{N}/N}(\mathcal{Q}_4)
    \otimes
    \left(
        \mathcal{A}^{(N,r)}_{\mathrm{BF}}
        \otimes
        \mathcal{A}^{(N,s)}_{\mathrm{BF}}
    \right)
    \nonumber\\
    \cong{}&
     \mathcal{T}^{(N)}_{\mathrm{BF}}(\mathcal Q_4)
    \otimes
    \mathcal{D}^{(10),\widetilde F}_{[r+s]_{N}/N}(\mathcal{Q}_4).
\end{align}
The object \( \mathcal{T}^{(N)}_{\mathrm{BF}}(\mathcal Q_4)\) denotes the residual topological sector
left by the fusion of the two auxiliary \(BF\)-type theories. In the local presentation
obtained by separating the diagonal and relative auxiliary fields, it can be written as
\begin{align}\label{ECUACION: DEFECTO SETOR ELECTRICO EN DIEZ DIMENSIONES}
    \mathcal{T}^{(N)}_{\mathrm{BF}}(\mathcal Q_4)
    =
    \int
    \mathscr{D}[\widehat{\Gamma}_{[2]}]\,
    \mathscr{D}[\widehat{\Gamma}_{[1]}]\,
    \exp\!\left[
        2\pi i\,N
        \int_{\mathcal{Q}_4}
        \widehat{\Gamma}_{[2]}\wedge d\widehat{\Gamma}_{[1]}
    \right].
\end{align}
This expression should be understood as a local differential-form presentation of the
residual sector. Here \(\widehat{\Gamma}_{[1]}\) and \(\widehat{\Gamma}_{[2]}\) are the
relative auxiliary fields that remain after extracting the diagonal \(BF\)-sector associated
with \(\mathcal{D}^{(10),\widetilde F}_{[r+s]_{N}/N}\). Thus,
\( \mathcal{T}^{(N)}_{\mathrm{BF}}(\mathcal Q_4)\) is a non-trivial topological sector supported on
\(\mathcal Q_4\), inherited from the fusion of the compactified auxiliary \(TQFT\)s.
Therefore, the effective group-like rule written above is only a shorthand. The genuine
fusion in the defect category is not purely group-like, because it contains this additional
topological remnant. More generally, for rational labels
\begin{align}
    \mathcal{D}^{(10),\widetilde F}_{1/N}(\mathcal{Q}_4)
    \otimes
    \mathcal{D}^{(10),\widetilde F}_{1/N}(\mathcal{Q}_4)
    \cong{}&
    \mathcal{T}^{(N)}_{\mathrm{BF}}(\mathcal{Q}_4)
    \otimes
    \mathcal{D}^{(10),\widetilde F}_{2/N}(\mathcal{Q}_4).
\end{align}
The non-invertibility is most sharply seen by fusing a defect with its dual, rather
than by considering \( \mathcal{D}\otimes\mathcal{D} \) alone. Indeed,
\begin{align}
    \mathcal{D}^{(10),\widetilde F}_{1/N}(\mathcal{Q}_4)
    \otimes
    \left(
        \mathcal{D}^{(10),\widetilde F}_{1/N}(\mathcal{Q}_4)
    \right)^{\vee}
    \neq
    \mathbf{1}.
\end{align}
\noindent
Thus, fusion with the adjoint does not return the tensor unit of the defect category.
Instead, it leaves a residual \(BF\)-type condensation sector. This is the categorical
signature that the compactified \(\widetilde F_{[4]}\)-defect is genuinely
non-invertible. Equivalently, the condensate of the compactified
twisted-\textit{Bianchi} sector is defined by
\begin{align}
    \mathcal{C}^{(10),\widetilde F}_{N}(\mathcal{Q}_4)
    &:=
    \mathcal{D}^{(10),\widetilde F}_{1/N}(\mathcal{Q}_4)
    \otimes
    \left(
        \mathcal{D}^{(10),\widetilde F}_{1/N}(\mathcal{Q}_4)
    \right)^{\dagger}
    \notag\\
    &\cong
    \int
    \mathscr{D}[\widehat{\Gamma}_{[1]}]\,
    \mathscr{D}[\widehat{\Gamma}_{[2]}]\,
    \exp\!\left[
        2\pi i N
        \int_{\mathcal{Q}_4}
        \widehat{\Gamma}_{[1]}\wedge
        \left(
            d\widehat{\Gamma}_{[2]}
            +
            \frac{H_{[3]}}{N}
        \right)
    \right]
    \notag\\
    &\qquad\otimes
    \int
    \mathscr{D}[\widehat{\Delta}_{[2]}]\,
    \mathscr{D}[\widehat{\Delta}_{[1]}]\,
    \exp\!\left[
        2\pi i N
        \int_{\mathcal{Q}_4}
        \widehat{\Delta}_{[2]}\wedge
        \left(
            d\widehat{\Delta}_{[1]}
            +
            \frac{F_{[2]}}{N}
        \right)
    \right].
\end{align}
Here \(\widehat{\Gamma}_{[1]}\) and \(\widehat{\Delta}_{[1]}\) are auxiliary compact
one-form gauge fields on \(\mathcal Q_4\), while
\(\widehat{\Gamma}_{[2]}\) and \(\widehat{\Delta}_{[2]}\) are auxiliary compact
two-form gauge fields. The first factor is the relative \(BF\)-type sector coupled to
the background \(H_{[3]}\), whereas the second factor is the diagonal \(BF\)-type
sector coupled to the \textit{Kaluza--Klein} curvature \(F_{[2]}\). This expression is understood
up to total derivatives on the defect support \(\mathcal Q_4\).
If the defect were invertible, this object would reduce to the identity. Instead, the
adjoint fusion leaves a residual \(BF\)-type topological sector, which is the categorical
signal of non-invertibility. This is the
categorical signal that the operator associated with the broken \textit{Bianchi} identity is
non-invertible. There is also a natural compatibility rule with the invertible \(H_{[3]}\)-sector. Recall that the vertical component of the eleven-dimensional \textit{Bianchi} operator produced the invertible Type IIA operator
\begin{align}
    U^{(10),H}_{\beta}(\mathcal{Q}_3)
    =
    \exp\!\left[
        2\pi i\,\beta
        \int_{\mathcal{Q}_3}
        H_{[3]}
    \right].
\end{align}
The compactified non-invertible defect absorbs only the subgroup of \(H_{[3]}\)-operators compatible
with the same \(N\)-stacking. Thus, for
\begin{align}
    \beta=\frac{k}{N},
    \qquad
    k\in\mathbb{Z},
\end{align}
and when the support of the invertible operator is contained in the support of the compactified
defect, one has the absorption rule
\begin{align}
    \mathcal{D}^{(10),\widetilde F}_{r/N}(\mathcal{Q}_4)
    \otimes
    U^{(10),H}_{k/N}(\mathcal{Q}_3)
    \simeq{}&
    \mathcal{D}^{(10),\widetilde F}_{r/N}(\mathcal{Q}_4),
    \\
        U^{(10),H}_{k/N}(\mathcal{Q}_3)
    \otimes
    \mathcal{D}^{(10),\widetilde F}_{r/N}(\mathcal{Q}_4)
    \simeq{}&
    \mathcal{D}^{(10),\widetilde F}_{r/N}(\mathcal{Q}_4)
\end{align}
with \(\mathcal{Q}_3\subset\mathcal{Q}_4\). The meaning is not that every invertible
\(H_{[3]}\)-operator is trivial. Rather, only the subgroup of the $H_{[3]}$ operators compatible with the same
\(N\)-stacking and with the topological constraint imposed by the defect is absorbed.

\subsection{The \textit{Bianchi} defect algebra after compactification}

\noindent
We now describe the fusion rules between the magnetic sectors after the
\textit{Kaluza--Klein} compactification, emphasizing the monoidal defect category of
SUGRA\((10,2A)\). In particular, we characterize its Picard group as the invertible
sector of the category.

\noindent
Let \(\mathfrak{C}^{10}_{\mathrm{Bianchi}}\) denote the monoidal topological defect category
generated by the compactified \textit{Bianchi} sector. Its invertible objects are the
\(H_{[3]}\)-defects \(U^{(10),H}_{\beta}\), while the twisted horizontal sector is generated
by the \(BF\)-dressed defects \(\mathcal{D}^{(10),\widetilde F}_{r/N}\). The tensor product in
\(\mathfrak{C}^{10}_{\mathrm{Bianchi}}\) is the fusion of topological defects, and isomorphisms
of objects will be denoted by \(\cong\).

\noindent
When the \(H_{[3]}\)-defect is moved through the support of the twisted
\(\widetilde F_{[4]}\)-defect, the corresponding topological relation is not an ordinary
commutation relation. It is better described by a crossing morphism
\begin{align}
    C_{\mathcal{D}_r,U_\beta}:
    \mathcal{D}^{(10),\widetilde F}_{r/N}(\mathcal{Q}_4)
    \otimes
    U^{(10),H}_{\beta}(\mathcal{Q}_3)
    \longrightarrow
    U^{(10),H}_{\beta}(\mathcal{Q}_3)
    \otimes
    \mathcal{D}^{(10),\widetilde F}_{r/N}(\mathcal{Q}_4).
\end{align}
In a sector where this morphism acts on a one-dimensional topological state space, it is
represented by the phase
\begin{align}
    C_{\mathcal{D}_r,U_\beta}
    =
    \exp\!\left[
        \frac{2\pi i}{N}\,
        r\,\beta\,
        \mathrm{Link}(\mathcal{Q}_4,\mathcal{Q}_3)
    \right].
\end{align}
Here
\(\mathrm{Link}(\mathcal{Q}_4,\mathcal{Q}_3)\in\mathbb{Z}\) denotes the
effective integer pairing selected by the auxiliary topological sector at the defect
junction. It should not be understood as an ordinary spacetime linking number in
\(M_{10}\), since
\begin{align}
    \dim(\mathcal Q_4)+\dim(\mathcal Q_3)=7\neq 9.
\end{align}
Rather, it denotes the topological junction pairing associated with the crossing move.

\noindent
Equivalently, after summing over the \(N\)-stacked \(H_{[3]}\)-sectors, one obtains the
condensation algebra
\begin{align}
    \mathcal{A}^{H}_{N}
    :=
    \bigoplus_{k\in\mathbb{Z}_{N}}
    U^{(10),H}_{k/N}.
\end{align}
Here \(\mathcal{A}^{H}_{N}\) is the algebra object obtained by summing the invertible
simple-current defects generated by \(U^{(10),H}_{1/N}\). The absorption rule is then most
naturally written in the \(\mathcal{A}^{H}_{N}\)-module category. The defect
\(\mathcal D^{(10),\widetilde F}_{r/N}\) is therefore regarded as an
\(\mathcal A^H_N\)-module object. After imposing the condensation relation, one may write
\begin{align}
    \mathcal{D}^{(10),\widetilde F}_{r/N}
    \otimes_{\mathcal A^{H}_{N}}
    \mathcal{A}^{H}_{N}
    &\cong
    \mathcal{D}^{(10),\widetilde F}_{r/N}.
\end{align}
In simple-current notation this becomes, for \(k\in\mathbb{Z}_{N}\),
\begin{align}
    \mathcal{D}^{(10),\widetilde F}_{r/N}
    \otimes
    U^{(10),H}_{k/N}
    \cong
    \mathcal{D}^{(10),\widetilde F}_{r/N},
\end{align}
with the understanding that this equivalence is taken in the condensed
\(\mathcal A^{H}_{N}\)-module category. Thus, the invertible \(H_{[3]}\)-operators
compatible with the same \(N\)-stacking \cite{Schafer-Nameki:2023jdn} are absorbed by
the non-invertible defect, whereas a general \(U^{(10),H}_{\beta}\) acts through the
corresponding crossing morphism.

\noindent
The compactified \textit{Bianchi} defect algebra can therefore be summarized as
\begin{align}
    U^{(10),H}_{\beta}
    \otimes
    U^{(10),H}_{\beta'}
    &\cong
    U^{(10),H}_{\beta+\beta'},
    \qquad
    \left(U^{(10),H}_{\beta}\right)^{\vee}
    \cong
    U^{(10),H}_{-\beta},
    \\
    \mathcal{D}^{(10),\widetilde F}_{r/N}
    \otimes
    \mathcal{D}^{(10),\widetilde F}_{s/N}
    &\cong
    \mathcal{T}^{(N)}_{\mathrm{BF}}
    \otimes
    \mathcal{D}^{(10),\widetilde F}_{[r+s]_{N}/N},
    \\
    \mathcal{D}^{(10),\widetilde F}_{r/N}
    \otimes
    \left(
        \mathcal{D}^{(10),\widetilde F}_{r/N}
    \right)^{\vee}
    &\cong
    \mathcal{C}^{(10),\widetilde F}_{r,N}
    \neq
    \mathbf{1},
    \\
    \mathcal{D}^{(10),\widetilde F}_{r/N}
    \otimes_{\mathcal A^{H}_{N}}
    \mathcal{A}^{H}_{N}
    &\cong
    \mathcal{D}^{(10),\widetilde F}_{r/N}.
\end{align}
Here \([r+s]_{N}\) denotes reduction modulo \(N\), and
\(\mathcal{T}^{(N)}_{\mathrm{BF}}\) is the residual \(BF\)-type topological sector
produced by fusing the auxiliary \(TQFT\)s carried by the two defects. The appearance of
this residual topological factor is the categorical signature of non-invertibility.
Consequently, the compactification of the eleven-dimensional \textit{Bianchi} sector does
not produce only an invertible Type IIA symmetry. It produces a mixed topological
symmetry category: an invertible \(H_{[3]}\)-sector and a twisted
\(\widetilde F_{[4]}\)-sector whose fusion is controlled by the compactified \(BF\)-type
topological theory.

\noindent
In categorical terms, the relevant monoidal defect category is
\begin{align}
    \mathfrak{C}^{10}_{\mathrm{Bianchi}}
    :=
    \left\langle
        \mathbf{1},
        \;
        U^{(10),H}_{\beta},
        \;
        \mathcal{D}^{(10),\widetilde F}_{r/N},
        \;
        \mathcal{T}^{(N)}_{\mathrm{BF}},
        \;
        \mathcal{C}^{(10),\widetilde F}_{r,N}
    \right\rangle_{\oplus,\otimes,\vee} .
\end{align}
Its invertible subcategory is
\begin{align}
    \left(\mathfrak{C}^{10}_{\mathrm{Bianchi}}\right)^{\times}
    =
    \left\langle
        U^{(10),H}_{\beta}
    \right\rangle_{\otimes},
\end{align}
and therefore its Picard group is the group of isomorphism classes of invertible objects,
\begin{align}
    \mathrm{Pic}
    \left(
        \mathfrak{C}^{10}_{\mathrm{Bianchi}}
    \right)
    :=
    \pi_{0}
    \left[
        \left(
            \mathfrak{C}^{10}_{\mathrm{Bianchi}}
        \right)^{\times}
    \right]
    \cong
    \left\{
        [U^{(10),H}_{\beta}]
        \;|\;
        \beta\in\mathbb{R}/\mathbb{Z}
    \right\}.
\end{align}
Here
\(\left(\mathfrak{C}^{10}_{\mathrm{Bianchi}}\right)^{\times}\)
denotes the Picard groupoid of
\(\mathfrak{C}^{10}_{\mathrm{Bianchi}}\), namely the maximal subgroupoid generated by
tensor-invertible defects and invertible morphisms. The symbol \(\pi_{0}\) denotes the set
of connected components of this groupoid. Equivalently, in the present defect category,
\begin{align}
    \pi_{0}
    \left[
        \left(\mathfrak{C}^{10}_{\mathrm{Bianchi}}\right)^{\times}
    \right]
\end{align}
is the set of isomorphism classes of invertible defects
\cite{Bhardwaj:2024xcx}, with group law induced by fusion
\cite{Shao:2023gho,Schafer-Nameki:2023jdn,Roumpedakis:2022aik}. If one restricts to
the \(N\)-stacked subcategory generated by \(U^{(10),H}_{1/N}\), this reduces to
\begin{align}
    \mathrm{Pic}_{N}
    \left(
        \mathfrak{C}^{10}_{\mathrm{Bianchi}}
    \right)
    \cong
    \mathbb{Z}_{N}.
\end{align}
The non-invertible sector is the monoidal subcategory generated by
\begin{align}
    \mathfrak{N}^{10}_{\widetilde F}
    :=
    \left\langle
        \mathcal{D}^{(10),\widetilde F}_{r/N},
        \;
        \mathcal{C}^{(10),\widetilde F}_{r,N},
        \;
        \mathcal{T}^{(N;r,s)}_{\mathrm{BF}}
    \right\rangle_{\otimes,(\cdot)^{\vee}}.
\end{align}
Thus, the invertible sector survives as the Picard part of the defect category, whereas
the twisted \(\widetilde F_{[4]}\)-sector is genuinely non-invertible because its adjoint
fusion does not return the tensor unit, but a residual \(BF\)-type topological sector.

\section{Charged probes and the M-theory/Type IIA dictionary}\label{SECCION: 6 Charged probes and the M-theory/Type IIA dictionary}

\noindent
The compactified defect algebra must be compatible with the standard
M-theory/Type IIA brane dictionary. Since the ten-dimensional defects arise as
\textit{Kaluza--Klein} descendants of eleven-dimensional operators, their direct charged
probes should be identified with the corresponding descendants of the M-theory branes.
Thus, the probe structure is inherited together with the defect structure, and provides a
check of the compactification at the level of defect actions, linking pairings and brane
charges
\cite{Witten:1995ex,Polchinski:1996na,Becker:2006dvp,Garcia-Valdecasas:2023mis}.
\begin{table}[h]
\centering
\small
\setlength{\tabcolsep}{4pt}
\renewcommand{\arraystretch}{1.15}
\begin{tabular}{c|c|c|c|c}
\hline
M-theory brane & Type IIA descendant & Sector & Defect support & Direct probe \\
\hline
\(M2_{\mathrm{unwrapped}}\)
& \(D2\)
& electric \(RR\)
& \(\Sigma_6\)
& \(W_{D2}\) \\
\(M5_{\mathrm{unwrapped}}\)
& \(NS5\)
& magnetic \(NSNS\)
& \(\mathcal Q_3\)
& \(W_{NS5}\) \\
\(M5_{\mathrm{wrapped}}\)
& \(D4\)
& twisted magnetic \(RR\)
& \(\mathcal Q_4\)
& \(W_{D4}\) \\
\hline
\end{tabular}
\caption{Direct charged probes inherited from the M-theory/Type IIA brane dictionary.}
\end{table}

\noindent
We now apply the linking criterion reviewed above. In ten dimensions, a defect supported
on a cycle \(S_k\) links a brane operator supported on a worldvolume \(W_\ell\) when
\begin{align}
    k+\ell=9.
\end{align}
Here \(\ell\) denotes the spacetime dimension of the brane worldvolume.\footnote{
We use the standard convention that a \(Dp\)-brane has worldvolume dimension \(p+1\).
See \cite{Polchinski:1996na,Becker:2006dvp}.
}
Therefore, the electric compactified defect
$\mathcal{D}^{(10),e}_{p/N}(\Sigma_6)$ directly links D2-brane operators, since
\begin{align}
    \dim(\Sigma_6)+\dim(W_{D2})=9.
\end{align}
This is the Type IIA descendant of the eleven-dimensional action on M2-brane operators,
\begin{align}
    M2_{\mathrm{unwrapped}}
    \longrightarrow
    D2.
\end{align}
Hence the D2 charge is not introduced independently in ten dimensions. It is inherited
from the compactification of the eleven-dimensional electric sector. This agrees with the
brane-probe interpretation of the Type IIA non-invertible defects in
\cite{Garcia-Valdecasas:2023mis}, where the dressed Supergravity defects act
non-trivially on the corresponding D-brane probes.

\noindent
The vertical component of the compactified \textit{Bianchi} sector gives the invertible
magnetic \(H_{[3]}\)-operator
\begin{align}
    U^{(10),H}_{\beta}(\mathcal Q_3)
    =
    \exp\!\left(
        2\pi i\,\beta
        \int_{\mathcal Q_3}H_{[3]}
    \right).
\end{align}
This operator links a six-dimensional worldvolume, hence its direct magnetic probe is
the \(NS5\)-brane:
\begin{align}
    \dim(\mathcal Q_3)+\dim(W_{NS5})=9.
\end{align}
Equivalently, in the presence of an \(NS5\)-source,
\begin{align}
    dH_{[3]}
    =
    q_{NS5}\,\delta^{(4)}(W_{NS5}),
\end{align}
so that
\begin{align}
    \int_{\mathcal Q_3}H_{[3]}
    =
    q_{NS5}\,
    \mathrm{Lk}(\mathcal Q_3,W_{NS5}).
\end{align}
Thus \(U^{(10),H}_{\beta}\) measures the magnetic \(NS5\)-charge. This is the vertical
component of the eleven-dimensional magnetic M5-sector, since the unwrapped M5-brane
descends to the \(NS5\)-brane,
\begin{align}
    M5_{\mathrm{unwrapped}}
    \longrightarrow
    NS5.
\end{align}

\noindent
The horizontal component gives the twisted magnetic \(RR\) sector. The corresponding
bare insertion would be
\begin{align}
    U^{(10),\widetilde F}_{\gamma}(\mathcal Q_4)
    =
    \exp\!\left(
        2\pi i\,\gamma
        \int_{\mathcal Q_4}\widetilde F_{[4]}
    \right),
\end{align}
but this operator is not topological in a generic \textit{Kaluza--Klein} fibration because
\begin{align}
    d\widetilde F_{[4]}
    =
    -
    H_{[3]}\wedge F_{[2]}.
\end{align}
Dimensionally, its direct magnetic probe is the D4-brane:
\begin{align}
    \dim(\mathcal Q_4)+\dim(W_{D4})=9.
\end{align}
In the presence of a D4-source, the twisted \textit{Bianchi} identity takes the schematic
form
\begin{align}
    d\widetilde F_{[4]}
    =
    -
    H_{[3]}\wedge F_{[2]}
    +
    q_{D4}\,\delta^{(5)}(W_{D4}).
\end{align}
Therefore, for \(\partial\mathcal V_5=\mathcal Q_4\),
\begin{align}
    \int_{\mathcal Q_4}\widetilde F_{[4]}
    =
    q_{D4}\,\mathrm{Lk}(\mathcal Q_4,W_{D4})
    -
    \int_{\mathcal V_5}H_{[3]}\wedge F_{[2]}.
\end{align}
This equation shows why the D4-brane is the direct magnetic probe of the
\(\widetilde F_{[4]}\)-sector, but also why the bare operator must be replaced by the
\(BF\)-dressed non-invertible defect
\(\mathcal D^{(10),\widetilde F}_{r/N}(\mathcal Q_4)\). This agrees with the Type IIA non-invertible defect found in \cite{Garcia-Valdecasas:2023mis}. The present virtue of the
construction is to explain its geometric origin as the horizontal compactified remnant of the
eleven-dimensional magnetic M5-sector:
\begin{align}
    M5_{\mathrm{wrapped}}
    \longrightarrow
    D4.
\end{align}
Thus, the charged probes obtained from the compactified defect algebra coincide with
the Type IIA brane probes found in \cite{Garcia-Valdecasas:2023mis}. The difference is
the interpretation: in the present construction the probe structure is not postulated
directly in ten dimensions, but follows from the \textit{Kaluza--Klein} fibration and the
standard M-theory/Type IIA brane dictionary. The eleven-dimensional M2 electric sector
descends to the D2 probe of the electric non-invertible defect, while the M5 magnetic
sector splits into the \(NS5\) probe of the vertical \(H_{[3]}\)-operator and the D4 probe
of the horizontal twisted \(\widetilde F_{[4]}\)-defect.

\section{Discussion and conclusions}\label{SECCION: 7 Discussion and conclusions}

\noindent
The analysis presented in this work shows that the defect structure of
eleven-dimensional Supergravity is transported non-trivially under
\textit{Kaluza--Klein} compactification. In the electric sector, the known
eleven-dimensional non-invertible defect associated with the Page-current obstruction
descends, in the zero-mode Supergravity regime, to a Type IIA non-invertible defect \cite{Garcia-Valdecasas:2023mis}.
This descent is not only a reduction of the fractional Page current. It also includes
the auxiliary topological sector required for gauge invariance:
\begin{align}
   \pi_* \mathcal D^{(11),e}_{q/N}(\Sigma_7)
    \cong
    \mathcal D^{(10),e}_{p/N}(\Sigma_6).
\end{align}
Thus, the Type IIA electric non-invertible defect is not an independent
ten-dimensional construction, but the compactified remnant of the corresponding
eleven-dimensional defect.

\noindent
The compactification of the \textit{Bianchi} sector gives a second, independent
effect. Although the eleven-dimensional identity
\begin{align}
    dF^{(11)}_{[4]}=0
\end{align}
defines an invertible magnetic HFS$(6)$, its \textit{Kaluza--Klein}
decomposition produces two different ten-dimensional sectors:
\begin{align}
    dH_{[3]}=0,
    \qquad
    d\widetilde F_{[4]}+H_{[3]}\wedge F_{[2]}=0 .
\end{align}
The first equation gives the ordinary invertible \(H_{[3]}\)-operator, while the second
one gives a twisted \(RR\) sector \cite{Becker:2006dvp}. Since \(\widetilde F_{[4]}\) is not closed by itself in
a generic \textit{Kaluza--Klein} background, the corresponding bare operator must be
dressed by the compactification of \(BF\)-type topological sector. Hence, an invertible
eleven-dimensional \textit{Bianchi} operator can give rise, after compactification, to
both an invertible and a non-invertible ten-dimensional sector.

\noindent
The resulting fusion rules show that the compactified defect algebra is not a group of
invertible operators. The fusion of two twisted \(\widetilde F_{[4]}\)-defects leaves a
residual \(BF\)-type topological sector,
\begin{align}
    \mathcal{D}^{(10),\widetilde F}_{r/N}
    \otimes
    \mathcal{D}^{(10),\widetilde F}_{s/N}
    \sim
    \mathcal{T}_{BF}^{(N)}
    \otimes
    \mathcal{D}^{(10),\widetilde F}_{[r+s]_N/N},
\end{align}
and the adjoint fusion produces a condensate rather than the tensor unit. Categorically,
the compactified \textit{Bianchi} sector is therefore a mixed monoidal defect category,
the invertible \(H_{[3]}\)-operators generate the Picard part \cite{Bhardwaj:2024xcx},
\begin{align}
    \mathrm{Pic}
    \left(
        \mathfrak C^{10}_{\mathrm{Bianchi}}
    \right)
    \cong
    \left\{
        [U^{(10),H}_{\beta}]
        \;|\;
        \beta\in\mathbb R/\mathbb Z
    \right\},
\end{align}
while the \(N\)-stacked restriction gives
\begin{align}
    \mathrm{Pic}_{N}
    \left(
        \mathfrak C^{10}_{\mathrm{Bianchi}}
    \right)
    \cong
    \mathbb Z_N.
\end{align}
The twisted \(\widetilde F_{[4]}\)-defects lie outside this Picard sector, since their
adjoint fusion leaves a non-trivial \(BF\)-type condensation sector.

\noindent
The charged probes give the corresponding physical interpretation. The electric
eleven-dimensional defect detects the unwrapped M2-brane sector, which descends to the
D2-brane probe of the Type IIA electric defect. The magnetic M5-brane sector splits
into the \(NS5\)-brane and \(D4\)-brane sectors. The \(NS5\) sector is associated with
the invertible \(H_{[3]}\)-operator, whereas the \(D4\) sector is governed by the twisted
\textit{Bianchi} identity for \(\widetilde F_{[4]}\). Thus, the same compactification that
maps M-branes into Type IIA branes also maps the corresponding charge-detecting
defects into the appropriate \(RR\) and \(NSNS\) defect sectors.

\noindent
The construction should be understood within the low-energy Supergravity approximation,
with the \textit{Kaluza--Klein} zero modes retained and the massive tower neglected. It
does not imply that full non-perturbative M-theory contains an exact global
non-invertible symmetry \cite{Harlow:2018tng}. Rather, the defects studied here are effective categorical
structures visible in the Supergravity regime. They encode how Page charges, large gauge
transformations \cite{Kalkkinen:2002tk}, flux quantization and brane charge conservation are reorganized by the
\textit{Kaluza--Klein} fibration \cite{Bergman:2004ne, Kaste:2002xs}. In this sense, Type IIA non-invertible defects are
compactified remnants of the eleven-dimensional M-theoretic defect structure.

\medskip
\textbf{Acknowledgements} \par 
FCP thanks to Physics area of the Chemistry Departament at U. Rioja, Spain and the Science Computation Research institute at the U. Rioja (SCRIUR), where part of this work was done and thanks to Instituto de Ciencias Exactas y Naturales (ICEN) at U. Arturo Prat, Iquique-Chile, where part of this work was done. FCP is supported by Doctorado nacional (ANID) 2023 Scholarship N$21230379$, project MATH-AMSUD 240048, and supported as graduate student in the “Doctorado en Física Mención Física-Matemática” Ph.D. program at the Universidad de Antofagasta. 
MPGM has been  partially supported by the PID2024-155685NB-C21 MCI Spanish Grants and by the University of La Rioja project REGI2025/41. AR  want to thank to MATH-AMSUD 240048  project and the Scientific Research Computing Institute of the University of La Rioja (SCRIUR), Spain. 

\appendix
\setcounter{equation}{0}
\renewcommand{\theequation}{A.\arabic{equation}}
\medskip
\bibliographystyle{MSP}
\bibliography{biblio}

@article{Gaiotto:2014kfa,
    author = "Gaiotto, Davide and Kapustin, Anton and Seiberg, Nathan and Willett, Brian",
    title = "{Generalized Global Symmetries}",
    eprint = "1412.5148",
    archivePrefix = "arXiv",
    primaryClass = "hep-th",
    doi = "10.1007/JHEP02(2015)172",
    journal = "JHEP",
    volume = "02",
    pages = "172",
    year = "2015"
}

@article{Garcia-Valdecasas:2023mis,
    author = "Garc{\'\i}a-Valdecasas, Eduardo",
    title = "{Non-invertible symmetries in supergravity}",
    eprint = "2301.00777",
    archivePrefix = "arXiv",
    primaryClass = "hep-th",
    doi = "10.1007/JHEP04(2023)102",
    journal = "JHEP",
    volume = "04",
    pages = "102",
    year = "2023"
}

@article{Fernandez-Melgarejo:2024ffg,
    author = "Fernandez-Melgarejo, Jose J. and Giorgi, Giacomo and Marques, Diego and Rosabal, J. A.",
    title = "{Noninvertible symmetries in type IIB supergravity}",
    eprint = "2407.09402",
    archivePrefix = "arXiv",
    primaryClass = "hep-th",
    doi = "10.1103/PhysRevD.111.066024",
    journal = "Phys. Rev. D",
    volume = "111",
    number = "6",
    pages = "066024",
    year = "2025"
}

@article{Heidenreich:2020pkc,
    author = "Heidenreich, Ben and McNamara, Jacob and Montero, Miguel and Reece, Matthew and Rudelius, Tom and Valenzuela, Irene",
    title = "{Chern-Weil global symmetries and how quantum gravity avoids them}",
    eprint = "2012.00009",
    archivePrefix = "arXiv",
    primaryClass = "hep-th",
    reportNumber = "ACFI-T20-16",
    doi = "10.1007/JHEP11(2021)053",
    journal = "JHEP",
    volume = "11",
    pages = "053",
    year = "2021"
}

@article{Cremmer:1978km,
    author = "Cremmer, E. and Julia, B. and Scherk, Joel",
    title = "{Supergravity Theory in 11 Dimensions}",
    reportNumber = "LPTENS-78-10",
    doi = "10.1016/0370-2693(78)90894-8",
    journal = "Phys. Lett. B",
    volume = "76",
    pages = "409--412",
    year = "1978"
}

@article{Montero:2017yja,
    author = "Montero, Miguel and Uranga, Angel M. and Valenzuela, Irene",
    title = "{A Chern-Simons Pandemic}",
    eprint = "1702.06147",
    archivePrefix = "arXiv",
    primaryClass = "hep-th",
    reportNumber = "IFT-UAM-CSIC-17-012, MPP-2017-26",
    doi = "10.1007/JHEP07(2017)123",
    journal = "JHEP",
    volume = "07",
    pages = "123",
    year = "2017"
}

@article{Schafer-Nameki:2023jdn,
    author = "Schafer-Nameki, Sakura",
    title = "{ICTP lectures on (non-)invertible generalized symmetries}",
    eprint = "2305.18296",
    archivePrefix = "arXiv",
    primaryClass = "hep-th",
    doi = "10.1016/j.physrep.2024.01.007",
    journal = "Phys. Rept.",
    volume = "1063",
    pages = "1--55",
    year = "2024"
}

@article{Gagliano:2024off,
    author = "Gagliano, Finn and Garc{\'\i}a Etxebarria, I{\~n}aki",
    title = "{SymTFTs for $U(1)$ symmetries from descent}",
    eprint = "2411.15126",
    archivePrefix = "arXiv",
    primaryClass = "hep-th",
    month = "11",
    year = "2024"
}

@article{Heckman:2024obe,
    author = "Heckman, Jonathan J. and McNamara, Jacob and Montero, Miguel and Sharon, Adar and Vafa, Cumrun and Valenzuela, Irene",
    title = "{Fate of stringy noninvertible symmetries}",
    eprint = "2402.00118",
    archivePrefix = "arXiv",
    primaryClass = "hep-th",
    reportNumber = "CERN-TH-2024-019",
    doi = "10.1103/PhysRevD.110.106001",
    journal = "Phys. Rev. D",
    volume = "110",
    number = "10",
    pages = "106001",
    year = "2024"
}

@article{Bhardwaj:2023kri,
    author = "Bhardwaj, Lakshya and Bottini, Lea E. and Fraser-Taliente, Ludovic and Gladden, Liam and Gould, Dewi S. W. and Platschorre, Arthur and Tillim, Hannah",
    title = "{Lectures on generalized symmetries}",
    eprint = "2307.07547",
    archivePrefix = "arXiv",
    primaryClass = "hep-th",
    doi = "10.1016/j.physrep.2023.11.002",
    journal = "Phys. Rept.",
    volume = "1051",
    pages = "1--87",
    year = "2024"
}

@inproceedings{Cordova:2022ruw,
    author = "Cordova, Clay and Dumitrescu, Thomas T. and Intriligator, Kenneth and Shao, Shu-Heng",
    title = "{Snowmass White Paper: Generalized Symmetries in Quantum Field Theory and Beyond}",
    booktitle = "{Snowmass 2021}",
    eprint = "2205.09545",
    archivePrefix = "arXiv",
    primaryClass = "hep-th",
    month = "5",
    year = "2022"
}

@article{GarciaEtxebarria:2022vzq,
    author = "Garc{\'\i}a Etxebarria, I{\~n}aki",
    title = "{Branes and Non-Invertible Symmetries}",
    eprint = "2208.07508",
    archivePrefix = "arXiv",
    primaryClass = "hep-th",
    doi = "10.1002/prop.202200154",
    journal = "Fortsch. Phys.",
    volume = "70",
    number = "11",
    pages = "2200154",
    year = "2022"
}

@article{Thorngren:2019iar,
    author = "Thorngren, Ryan and Wang, Yifan",
    title = "{Fusion category symmetry. Part I. Anomaly in-flow and gapped phases}",
    eprint = "1912.02817",
    archivePrefix = "arXiv",
    primaryClass = "hep-th",
    reportNumber = "PUPT-2603",
    doi = "10.1007/JHEP04(2024)132",
    journal = "JHEP",
    volume = "04",
    pages = "132",
    year = "2024"
}

@inproceedings{Shao:2023gho,
    author = "Shao, Shu-Heng",
    title = "{What's Done Cannot Be Undone: TASI Lectures on Non-Invertible Symmetries}",
    booktitle = "{Theoretical Advanced Study Institute in Elementary Particle Physics 2023}: {Aspects of Symmetry}",
    eprint = "2308.00747",
    archivePrefix = "arXiv",
    primaryClass = "hep-th",
    reportNumber = "YITP-SB-2023-19",
    month = "8",
    year = "2023"
}

@article{Frohlich:2004ef,
    author = "Frohlich, Jurg and Fuchs, Jurgen and Runkel, Ingo and Schweigert, Christoph",
    title = "{Kramers-Wannier duality from conformal defects}",
    eprint = "cond-mat/0404051",
    archivePrefix = "arXiv",
    reportNumber = "HU-EP-04-19",
    doi = "10.1103/PhysRevLett.93.070601",
    journal = "Phys. Rev. Lett.",
    volume = "93",
    pages = "070601",
    year = "2004"
}

@article{Ishii:2007sy,
    author = "Ishii, Takaaki and Ishiki, Goro and Ohta, Kazutoshi and Shimasaki, Shinji and Tsuchiya, Asato",
    title = "{On relationships among Chern-Simons theory, BF theory and matrix model}",
    eprint = "0711.4235",
    archivePrefix = "arXiv",
    primaryClass = "hep-th",
    reportNumber = "OU-HET-592, TU-786",
    doi = "10.1143/PTP.119.863",
    journal = "Prog. Theor. Phys.",
    volume = "119",
    pages = "863--882",
    year = "2008"
}

@article{Kalkkinen:2002tk,
    author = "Kalkkinen, Jussi and Stelle, K. S.",
    title = "{Large gauge transformations in M theory}",
    eprint = "hep-th/0212081",
    archivePrefix = "arXiv",
    reportNumber = "IMPERIAL-TP-01-02-25, IHES-P-02-57",
    doi = "10.1016/S0393-0440(03)00027-5",
    journal = "J. Geom. Phys.",
    volume = "48",
    pages = "100--132",
    year = "2003"
}

@inproceedings{Marolf:2000cb,
    author = "Marolf, Donald",
    title = "{Chern-Simons terms and the three notions of charge}",
    booktitle = "{International Conference on Quantization, Gauge Theory, and Strings: Conference Dedicated to the Memory of Professor Efim Fradkin}",
    eprint = "hep-th/0006117",
    archivePrefix = "arXiv",
    reportNumber = "SUGP-00-6-1",
    pages = "312--320",
    month = "6",
    year = "2000"
}

@article{Mickler:2015eca,
    author = "Mickler, Ryan",
    title = "{Localization for Chern{\textendash}Simons on circle bundles via loop groups}",
    eprint = "1507.01626",
    archivePrefix = "arXiv",
    primaryClass = "math.DG",
    doi = "10.1016/j.geomphys.2018.06.005",
    journal = "J. Geom. Phys.",
    volume = "132",
    pages = "257--281",
    year = "2018"
}

@article{Witten:1995ex,
    author = "Witten, Edward",
    title = "{String theory dynamics in various dimensions}",
    eprint = "hep-th/9503124",
    archivePrefix = "arXiv",
    reportNumber = "IASSNS-HEP-95-18",
    doi = "10.1016/0550-3213(95)00158-O",
    journal = "Nucl. Phys. B",
    volume = "443",
    pages = "85--126",
    year = "1995"
}

@book{Becker:2006dvp,
    author = "Becker, K. and Becker, M. and Schwarz, J. H.",
    title = "{String theory and M-theory: A modern introduction}",
    doi = "10.1017/CBO9780511816086",
    isbn = "978-0-511-25486-4, 978-0-521-86069-7, 978-0-511-81608-6",
    publisher = "Cambridge University Press",
    month = "12",
    year = "2006"
}

@article{Horava:1995qa,
    author = "Horava, Petr and Witten, Edward",
    title = "{Heterotic and Type I string dynamics from eleven dimensions}",
    eprint = "hep-th/9510209",
    archivePrefix = "arXiv",
    reportNumber = "IASSNS-HEP-95-86, PUPT-1571A",
    doi = "10.1016/0550-3213(95)00621-4",
    journal = "Nucl. Phys. B",
    volume = "460",
    pages = "506--524",
    year = "1996"
}

@article{Cordova:2022ieu,
    author = "Cordova, Clay and Ohmori, Kantaro",
    title = "{Noninvertible Chiral Symmetry and Exponential Hierarchies}",
    eprint = "2205.06243",
    archivePrefix = "arXiv",
    primaryClass = "hep-th",
    doi = "10.1103/PhysRevX.13.011034",
    journal = "Phys. Rev. X",
    volume = "13",
    number = "1",
    pages = "011034",
    year = "2023"
}

@article{Choi:2022jqy,
    author = "Choi, Yichul and Lam, Ho Tat and Shao, Shu-Heng",
    title = "{Noninvertible Global Symmetries in the Standard Model}",
    eprint = "2205.05086",
    archivePrefix = "arXiv",
    primaryClass = "hep-th",
    reportNumber = "YITP-SB-2022-21, MIT-CTP/5433",
    doi = "10.1103/PhysRevLett.129.161601",
    journal = "Phys. Rev. Lett.",
    volume = "129",
    number = "16",
    pages = "161601",
    year = "2022"
}

@article{Choi:2022fgx,
    author = "Choi, Yichul and Lam, Ho Tat and Shao, Shu-Heng",
    title = "{Non-invertible Gauss law and axions}",
    eprint = "2212.04499",
    archivePrefix = "arXiv",
    primaryClass = "hep-th",
    reportNumber = "MIT-CTP/5504, YITP-SB-2022-39",
    doi = "10.1007/JHEP09(2023)067",
    journal = "JHEP",
    volume = "09",
    pages = "067",
    year = "2023"
}

@article{Bergshoeff:1996ui,
    author = "Bergshoeff, E. and de Roo, M. and Green, Michael B. and Papadopoulos, G. and Townsend, P. K.",
    title = "{Duality of type II 7 branes and 8 branes}",
    eprint = "hep-th/9601150",
    archivePrefix = "arXiv",
    reportNumber = "DAMTP-R-95-55-REV, UG-15-95",
    doi = "10.1016/0550-3213(96)00171-X",
    journal = "Nucl. Phys. B",
    volume = "470",
    pages = "113--135",
    year = "1996"
}

@article{Witten:1998cd,
    author = "Witten, Edward",
    title = "{D-branes and K-theory}",
    eprint = "hep-th/9810188",
    archivePrefix = "arXiv",
    reportNumber = "IASSNS-HEP-98-82",
    doi = "10.1088/1126-6708/1998/12/019",
    journal = "JHEP",
    volume = "12",
    pages = "019",
    year = "1998"
}

@article{Minasian:1997mm,
    author = "Minasian, Ruben and Moore, Gregory W.",
    title = "{K theory and Ramond-Ramond charge}",
    eprint = "hep-th/9710230",
    archivePrefix = "arXiv",
    reportNumber = "YCTP-P21-97",
    doi = "10.1088/1126-6708/1997/11/002",
    journal = "JHEP",
    volume = "11",
    pages = "002",
    year = "1997"
}

@article{Banks:2010zn,
    author = "Banks, Tom and Seiberg, Nathan",
    title = "{Symmetries and Strings in Field Theory and Gravity}",
    eprint = "1011.5120",
    archivePrefix = "arXiv",
    primaryClass = "hep-th",
    doi = "10.1103/PhysRevD.83.084019",
    journal = "Phys. Rev. D",
    volume = "83",
    pages = "084019",
    year = "2011"
}

@article{Harlow:2018tng,
    author = "Harlow, Daniel and Ooguri, Hirosi",
    title = "{Symmetries in quantum field theory and quantum gravity}",
    eprint = "1810.05338",
    archivePrefix = "arXiv",
    primaryClass = "hep-th",
    doi = "10.1007/s00220-021-04040-y",
    journal = "Commun. Math. Phys.",
    volume = "383",
    number = "3",
    pages = "1669--1804",
    year = "2021"
}

@book {MR0152974,
    AUTHOR = {Kobayashi, Shoshichi and Nomizu, Katsumi},
     TITLE = {Foundations of differential geometry. {V}ol {I}},
 PUBLISHER = {Interscience Publishers, a division of John Wiley \& Sons, New
              York-Lond on},
      YEAR = {1963},
     PAGES = {xi+329},
   MRCLASS = {53.00},
  MRNUMBER = {0152974 (27 \#2945)},
MRREVIEWER = {J. Eells},
  BOEKCODE = {53-xx},
}

@article{Duff:1996aw,
    author = "Duff, M. J.",
    title = "{M Theory (The Theory Formerly Known as Strings)}",
    eprint = "hep-th/9608117",
    archivePrefix = "arXiv",
    reportNumber = "CTP-TAMU-33-96",
    doi = "10.1142/S0217751X96002583",
    journal = "Int. J. Mod. Phys. A",
    volume = "11",
    pages = "5623--5642",
    year = "1996"
}

@article{Choi:2022zal,
    author = "Choi, Yichul and Cordova, Clay and Hsin, Po-Shen and Lam, Ho Tat and Shao, Shu-Heng",
    title = "{Non-invertible Condensation, Duality, and Triality Defects in 3+1 Dimensions}",
    eprint = "2204.09025",
    archivePrefix = "arXiv",
    primaryClass = "hep-th",
    reportNumber = "YITP-SB-2022-16, MIT/CTP-5423, YITP-SB-2022-16, MIT/CTP-5423",
    doi = "10.1007/s00220-023-04727-4",
    journal = "Commun. Math. Phys.",
    volume = "402",
    number = "1",
    pages = "489--542",
    year = "2023"
}

@article{Roumpedakis:2022aik,
    author = "Roumpedakis, Konstantinos and Seifnashri, Sahand and Shao, Shu-Heng",
    title = "{Higher Gauging and Non-invertible Condensation Defects}",
    eprint = "2204.02407",
    archivePrefix = "arXiv",
    primaryClass = "hep-th",
    reportNumber = "YITP-SB-2022-14",
    doi = "10.1007/s00220-023-04706-9",
    journal = "Commun. Math. Phys.",
    volume = "401",
    number = "3",
    pages = "3043--3107",
    year = "2023"
}

@article{Apruzzi:2023uma,
    author = "Apruzzi, Fabio and Bonetti, Federico and Gould, Dewi S. W. and Schafer-Nameki, Sakura",
    title = "{Aspects of categorical symmetries from branes: SymTFTs and generalized charges}",
    eprint = "2306.16405",
    archivePrefix = "arXiv",
    primaryClass = "hep-th",
    doi = "10.21468/SciPostPhys.17.1.025",
    journal = "SciPost Phys.",
    volume = "17",
    number = "1",
    pages = "025",
    year = "2024"
}

@article{Kaste:2002xs,
    author = "Kaste, Peter and Minasian, Ruben and Petrini, Michela and Tomasiello, Alessandro",
    title = "{Kaluza-Klein bundles and manifolds of exceptional holonomy}",
    eprint = "hep-th/0206213",
    archivePrefix = "arXiv",
    reportNumber = "CPHT-RR-039-0502",
    doi = "10.1088/1126-6708/2002/09/033",
    journal = "JHEP",
    volume = "09",
    pages = "033",
    year = "2002"
}

@article{Bergman:2004ne,
    author = "Bergman, Aaron and Varadarajan, Uday",
    title = "{Loop groups, Kaluza-Klein reduction and M-theory}",
    eprint = "hep-th/0406218",
    archivePrefix = "arXiv",
    reportNumber = "UTTG-06-04",
    doi = "10.1088/1126-6708/2005/06/043",
    journal = "JHEP",
    volume = "06",
    pages = "043",
    year = "2005"
}

@inproceedings{Polchinski:1996na,
    author = "Polchinski, Joseph",
    title = "{Tasi lectures on D-branes}",
    booktitle = "{Theoretical Advanced Study Institute in Elementary Particle Physics (TASI 96): Fields, Strings, and Duality}",
    eprint = "hep-th/9611050",
    archivePrefix = "arXiv",
    reportNumber = "NSF-ITP-96-145",
    pages = "293--356",
    month = "11",
    year = "1996"
}

@article{Bhardwaj:2024xcx,
    author = "Bhardwaj, Lakshya and D{\'e}coppet, Thibault and Schafer-Nameki, Sakura and Yu, Matthew",
    title = "{Fusion 3-Categories for Duality Defects}",
    eprint = "2408.13302",
    archivePrefix = "arXiv",
    primaryClass = "math.CT",
    doi = "10.1007/s00220-025-05388-1",
    journal = "Commun. Math. Phys.",
    volume = "406",
    number = "9",
    pages = "208",
    year = "2025"
}

@article{Brennan:2024fgj,
    author = "Brennan, T. Daniel and Sun, Zhengdi",
    title = "{A SymTFT for continuous symmetries}",
    eprint = "2401.06128",
    archivePrefix = "arXiv",
    primaryClass = "hep-th",
    doi = "10.1007/JHEP12(2024)100",
    journal = "JHEP",
    volume = "12",
    pages = "100",
    year = "2024"
}
\end{document}